\documentclass[article]{JHEP3}

\usepackage{amsmath,epsfig}

\usepackage{amssymb,amsfonts}

\title{Closed Strings in Misner Space:\\
Stringy Fuzziness with a Twist}

\preprint{\hepth{0407216}\\LPTHE-04-19\\LPTENS-04-34\\WIS/19/04-JUL-DPP}

\author{M.~Berkooz$^\diamondsuit$\footnote{Incumbent of the Recanati career
development chair for energy research}~,
B.~Durin$^\clubsuit$, B.~Pioline$^{\clubsuit\spadesuit}$,
D.~Reichmann$^\diamondsuit$\\\\
$^\diamondsuit$~Weizmann Institute of Science,\\
Rehovot 76100, Israel\\ \\
$\clubsuit$~LPTHE, Universit\'es Paris 6 et 7, 4 place Jussieu, \\
75252 Paris cedex 05, France
\\ \\
$\spadesuit$~LPTENS, D\'epartement de Physique de l'ENS, 24 rue Lhomond,\\
75231 Paris cedex 05, France
\\ \\
{\tt E-mail:\\
berkooz@wisemail.weizmann.ac.il,
bdurin@lpthe.jussieu.fr,
pioline@lpthe.jussieu.fr,
tragula@wisemail.weizmann.ac.il}
}

\abstract{Misner space, also known as the Lorentzian orbifold
$\Real^{1,1}/boost$, is the simplest tree-level solution of string theory
with a cosmological singularity. We compute tree-level scattering amplitudes
involving twisted states, using operator and current algebra techniques.
We find that, due to zero-point quantum fluctuations of the excited modes,
twisted strings with a large winding number $w$ are fuzzy on
a scale $\sqrt{\log w}$, which can be much larger than the string
scale. Wave functions are smeared by an operator
$\exp\left( \Delta(\nu) \pa_+ \pa_-\right)$ reminiscent of
the Moyal-product of non-commutative geometry, which, since $\Delta(\nu)$ is
real, modulates the amplitude rather than the phase of the wave function,
and is purely gravitational in its origin.
We compute the scattering amplitude of two twisted states and one
tachyon or graviton, and find a finite result. The scattering amplitude of
two twisted and two untwisted states is found to diverge, due to the
propagation of intermediate winding strings with vanishing boost momentum.
The scattering amplitude of three twisted fields is computed
by analytic continuation from three-point amplitudes of states with
non-zero $p^+$
in the Nappi-Witten plane wave, and the non-locality of the
three-point vertex is found to diverge for
certain kinematical configurations. Our results for the three-point
amplitudes allow in principle to compute, to leading order,
the back-reaction on the
metric due to a condensation of coherent winding strings.
}

\makeatletter
\renewcommand{\subsubsection}{\@startsection{subsubsection}{3}{0mm}{-\baselineskip}{0.5\baselineskip}{\normalfont\normalsize\it}}
\makeatother

\newcommand{\pa}{\partial}
\newcommand{\xp}{{x^+}}
\newcommand{\xm}{{x^-}}

\newcommand{\p}{\partial}
\newcommand{\s}{\sigma}
\newcommand{\nn}{\nonumber}
\newcommand{\eps}{\epsilon}
\newcommand{\Real}{\mathbb{R}}

\def\bea{\begin{eqnarray}}
\def\eea{\end{eqnarray}}
\def\be{\begin{equation}}
\def\ee{\end{equation}}
\def\ba{\begin{align}}
\def\ea{\end{align}}
\def\bse{\begin{subequations}}
\def\ese{\end{subequations}}
\def\bi{\begin{itemize}}
\def\ei{\end{itemize}}

\def\a{\alpha}
\def\talpha{\tilde\alpha}

\def\teps{\tilde\epsilon}
\def\tzeta{\tilde\zeta}
\def\tDelta{\tilde\Delta}

\def\ta{\tilde \alpha}
\def\tM{\tilde M}
\def\1F1{{}_1\!F_1}
\def\2F0{{}_2\!F_0}

\newcommand{\abs}[1]{\left\vert#1\right\vert}

\newcommand{\cno}[1]{\,\textbf{:}#1\textbf{:}\,}
\newcommand{\vecs}[1]{\vec{#1}^{\;2}}
\newcommand{\vecd}[2]{\vec{#1}\cdot\vec{#2}}
\newcommand{\Rf}{\mathbb{R}}    \newcommand{\Zf}{\mathbb{Z}}
       
\newcommand{\A}{{\cal{A}}}      \newcommand{\V}{{\cal V}}
         
       \newcommand{\at}{{\alpha'}}
\newcommand{\tp}{(2\pi)}        
\newcommand{\csp}[1]{\hspace{#1em},\hspace{#1em}}
\newcommand{\hsp}[1]{\hspace{#1em}}
\newcommand{\rar}[1]{\hspace{#1em}\Rightarrow\hspace{#1em}}
\newcommand{\epp}{{\epsilon^+}} \newcommand{\epm}{{\epsilon^-}}
\newcommand{\ppp}{{p^+}}        \newcommand{\ppm}{{p^-}}
\newcommand{\Xp}{{X^+}}         \newcommand{\Xm}{{X^-}}
 
\DeclareMathOperator{\csch}{csch}

\begin{document}
\maketitle \setcounter{tocdepth}{2}
\tableofcontents

\section{Introduction}

Misner space \cite{Misner},
also known as the Lorentzian orbifold $\mathbb{R}^{1,1}/boost$,
is an especially attractive exact solution of string theory at
tree-level \cite{Horowitz:ap}.
Since it involves two cosmological (Milne) regions connected by a space-like
singularity, it is an interesting toy model to study whether quantum effects
may resolve cosmological singularities, and give ground for pre-Big Bang
scenarios \cite{Seiberg:2002hr} (see \cite{Nappi:1992kv,Elitzur:2002rt,
lms,lms2,
Craps:2002ii,Balasubramanian:2002ry,Johnson:2004zq}
for other $\alpha'$-exact models of
cosmological singularities).
Since it also involves two static (Rindler) regions with compact time
attached at the singularity, it may also be a useful setting to discuss
chronology protection in string theory. Finally, Misner's cousin
Grant space \cite{Grant:1992kj},
the orbifold $\mathbb{R}^{1,2}/(boost\times translation)$,
arises as a limit of the inside of BTZ black hole \cite{btz,Cornalba:2002fi},
hence may have bearing on the fate of certain black hole singularities.

Being a quotient of flat space, Misner space may be amenable to
standard conformal theory techniques, properly extended to account
for its Lorentzian signatures and  non-compact nature \cite{Nekrasov:2002kf}. 
In particular, it includes both untwisted states, 
corresponding to gravitons and
other particles propagating in the geometry, and an infinite number of twisted
sectors, corresponding to strings winding around the compact, space-like
or time-like, directions \cite{Pioline:2003bs,Berkooz:2004re}.
While tree-level scattering amplitudes
of untwisted states have been shown to lead to
severe divergences  \cite{Berkooz:2002je,Horowitz:2002mw}
(see \cite{lms,lms2} for a study of scattering amplitudes in a
related model), we have shown in \cite{Pioline:2003bs, Berkooz:2004re}
that twisted strings, both in the
Milne and Rindler regions, are pair-produced quantum-mechanically, and found
indications that the resulting back-reaction may in fact resolve
the singularity. While a consistent treatment of the back-reaction of
correlated pairs of winding strings is still out of reach, a somewhat less
ambitious approach is to marginally deform the orbifold CFT by
a coherent superposition of twisted strings with
a given winding number, and determine the
deformed metric order by order in the conformal perturbation, if
the true conformal field theory happens to be close enough from 
the orbifold point.

In this paper, we take another step in this direction, and study
tree-level scattering of twisted and
untwisted sector states of the bosonic string in Misner space.
Amplitudes involving two twisted states can
be computed by operator techniques, after expressing untwisted vertex
operators in terms of the mode expansion in the twisted sector with
winding number $w$. We carry this out in Section 2 below, and find that, due
to the zero-point quantum fluctuations of the excited modes,
the energy momentum
and other untwisted sector currents sourced by a twisted string, are
smeared over a characteristic distance $\sqrt{\Delta(\nu)}$,
where $\Delta(\nu)$, displayed in \eqref{exctac} below, is the regulated
r.m.s. size of the string  in the twisted sector
$w = - \nu/\beta$, where $\beta$ is the rapidity of the boost transformation.
In particular, for large winding number $w$, the size grows as
$\Delta(\nu) \sim \log w$, and can thus become arbitrarily larger than
the string scale $\alpha'$. This fuzziness is qualitatively similar
to non-commutative geometry, in as much as sources are smeared by
an exponential of the square of the momentum, $\exp\left[ -p^+ p^- \Delta(\nu)
\right]$. The extended nature of the string can therefore manifest itself down
to very low energies if $\Delta(\nu)$ is large enough. 
In contrast to the Moyal product however, this non-locality
arises at the level of charged states rather than dipoles, and modulates
the amplitude rather than the phase of the wave function.  It arises
purely in the closed string sector, through gravitational effects due to
cosmological singularity. This non-locality implies that a 
condensation of long strings
in the Rindler regions may have an important effect in the cosmological
regions, and possibly resolve the singularity. It is tempting to speculate
that such a non-locality may also be relevant for the dynamics of
the BTZ black hole inside the horizon.

In Section 3, using this twisted mode representation for the
untwisted vertex operators, we compute three-point amplitudes involving
two twisted states in their
``tachyonic'' ground state (a non-vanishing transverse momentum can be chosen
to ensure that the state is in fact not tachyonic in the two light-cone
directions) and one off-shell tachyon or graviton.
The result can be interpreted
as the tachyon source or energy-momentum sourced by a single twisted string.
Using a real space representation for the ``quasi-zero-modes'' of the string,
and the relation to the charged particle problem discussed
in \cite{Pioline:2003bs, Berkooz:2004re}, we find that the amplitude is given
by the overlap of the wave functions of two charged particles and one neutral
particle, up to the smearing kernel discussed above. It is in particular
finite. We proceed to study the 4-point amplitude of two twisted states
and two untwisted ones (two tachyons, or one tachyon and one graviton),
and find that the amplitude diverges whenever the intermediate winding
string has zero boost-momentum: this appears to be related to a large
emission of winding strings close to the cosmological singularity.
This result parallels the discussion of the four-untwisted scattering
amplitude in \cite{Berkooz:2002je}, which was found to diverge
due to large graviton exchange near the singularity.

Finally, in Section 4, we turn to amplitudes involving more than two
twisted states: in this case, operator methods are no longer applicable.
Nevertheless, correlators of Lorentzian twist fields can be obtained
by analytic continuation from those of Euclidean twist fields, i.e.
fields creating a cut
$(Z \to e^{2\pi i\theta} Z, \bar Z \to e^{-2\pi i\theta} \bar Z)$
on the Euclidean two-plane $(Z,\bar Z)$, with an generally
irrational angle $\theta$.
Such twist fields occur as subfactors
in the vertex operators of states with $p^+=\theta\neq 0$ in the Nappi-Witten
plane wave,
and their correlation functions have been computed recently
using current algebra techniques \cite{Kiritsis:jk,D'Appollonio:2003dr}
or a Wakimoto-type representation \cite{Cheung:2003ym,Bianchi:2004vf}.
Using this correspondence, we compute the three-point vertex of three twisted
states, and find its real-space representation \eqref{ker3}. Again,
we find that the interaction vertex is spread over a distance
$\Xi(\nu_1,\nu_2)$ given in \eqref{xi}. For special kinematical
configurations, this non-locality scale becomes infinite: it would
be interesting to understand the precise physics behind this
singularity.

Using our results in Section 3 and 4, one may now compute the back-reaction on
the tachyon and massless, twisted or untwisted fields, induced by
the condensation of a single twisted string, to leading order in
the deformation parameter. 
We leave this analysis to further work.

For completeness, scattering amplitudes of untwisted states are
discussed in Appendix A, extending the analysis of \cite{Berkooz:2002je,lms,
lms2} to the cases of gravitons in the bosonic string / superstring in
Misner / Grant space. We find that, 
unlike the null brane, constructing wave-packets
with continuous boost momentum does not help in removing divergences in
Grant space, although it does reduce the range of dangerous momenta.


\section{Review of the first quantization of twisted strings}

\subsection{The geometry of the orbifold}

Recall that Misner space is defined as the orbifold of
two-dimensional flat Minkowski space $\Real^{1,1}$ by the discrete boost
$(x^+,x^-)\rightarrow (e^{2\pi\beta} x^+, e^{-2\pi\beta}x^-)$. The geometry
of the quotient consists of four Lorentzian cones touching at their
apex.  The regions $x^+<0,x^-<0$ (P) and $x^+>0,x^->0$ (F),
often called  Milne regions, are cosmological geometries with a
shrinking spatial circle collapsing in a Big Crunch, and an
expanding spatial circle which begins in a Big Bang, respectively. The
regions $x^+>0,x^-<0$ (R) and $x^+<0,x^->0$ (L), often termed Rindler
regions, are time-independent geometries with a compact time:
indeed, the Killing  vector generating the boost is space-like
in the Milne regions, but becomes time-like in
the Rindler regions.

As in regular Euclidean orbifolds, the spectrum consists of two kinds
of states - untwisted and twisted. Untwisted strings correspond to
 particle-like
excitations which are invariant under the orbifold action; for
non-tachyonic two-dimensional mass  $\mu^2$, these states are  mostly
localized in the Milne regions, although their wave function
extends  to a distance $r\sim |j|/\mu$ in the Rindler regions, where
the ``boost momentum'' $j$ is the quantized momentum
along the compact spatial circle (or energy with respect to the
compact Rindler time). The twisted states will be reviewed momentarily
after we introduce some of the required machinery.

Our goal in this paper is to study the scattering of untwisted
states off a twisted string, and identify the sources that the twisted modes create for the untwisted graviton and tachyon. A effect that we will discuss depends crucially on the a correct treatment of both the excited and the quasi-zero modes of the string, which we will now survey.

\subsection{Quantization of the twisted sectors}

Closed strings in the $w$-th twisted sector of the Lorentzian
satisfy the twisted periodicity condition around the cylinder,
\be
X^{\pm}(\sigma+2\pi,\tau)=e^{\pm 2\pi w \beta} X^{\pm}(\sigma,\tau)
\ee
The embedding coordinates $X^\pm$ thus admit the normal mode expansion
(in the units where $\alpha'=2$)
\be
\label{closedmod}
X^\pm(\tau,\sigma)= i \sum_{n=-\infty}^{\infty}
(n \pm i\nu)^{-1} \alpha_n^\pm e^{-i(n\pm i \nu)(\tau-\sigma)}
+ i \sum_{n=-\infty}^{\infty}
(n \mp i\nu)^{-1}
\tilde \alpha_n^\pm e^{-i(n\mp i \nu)(\tau+\sigma)}
\ee
where the imaginary part of the frequency is related to the
winding number and boost parameter by $\nu=-w \beta$.
The coordinates $\tau,\sigma$ denote the (non-compact) time and
(compact) spatial directions of the Lorentzian world-sheet,
and are related to the Euclidean coordinates of radial quantization
by $z=e^{i(\tau-\sigma)},\,\bar z=e^{i(\tau+\sigma)}$.

As for strings in flat space,  left and right-moving excitations
are decoupled, and satisfy the following
commutation relations and hermiticity properties,
\bea
[\alpha_m^+, \alpha_n^-] =
-(m+i\nu) \delta_{m+n}\ &,&\quad
[\tilde \alpha_m^+, \tilde \alpha_n^-] =
-(m-i\nu) \delta_{m+n}\ ,\quad
\\
(\alpha_{-n}^\pm)^* = \alpha_n^\pm\ &,&\quad (\tilde
\alpha_{-n}^\pm)^* = \tilde \alpha_n^\pm
\eea
Of particular interest are the (quasi) zero-modes $\alpha_0^\pm$
and $\ta_0^\pm$, which are self-adjoint and satisfy
\be
\label{0m}
[\a^+, \a^-]=-i\nu\ ,
\quad[\talpha^+, \talpha^-]=i\nu
\ee
where we dropped the subscript 0 for conciseness.
Excited modes may be quantized just as in flat
space, as creation and annihilation operators in a Fock space
with ground state annihilated by $\alpha^\pm_{m>0}$.
In contrast, due to their unusual hermiticity properties,
zero-modes require a particular treatment.
They may be represented as covariant derivatives
for a charged scalar field in an electric field \cite{Pioline:2003bs},
 \be
\label{reals} \a^\pm= i \nabla_{\mp} = i \p_\mp
\pm \frac{\nu}{2} x^\pm\ ,\quad \talpha^\pm= i
\tilde \nabla_{\mp} = i \p_\mp \mp \frac{\nu}{2}
x^\pm
\ee
which ensures the proper hermiticity properties under
the inner product
\be
\label{hermi}
\langle f_1 | f_2 \rangle = \int dx^+ dx^- ~
f_1^*(x^+,x^-)~f_2(x^+,x^-)
\ee
On-shell wave functions satisfying the physical
state conditions $L_0=\bar L_0=1$ are thus eigenmodes
of the Klein-Gordon operator for a charged particle in an electric field,
\be
\label{vir}
\left(\alpha_0^+ \alpha_0^- + \alpha_0^- \alpha_0^+ \right) f = M^2 f \
,\quad
\left(\talpha_0^+ \talpha_0^-  + \talpha_0^- \talpha_0^+ \right) f =
\tilde M^2 f
\ee
where $M^2$ and $\tM^2$ are the contributions from the left and
right moving excited modes. The orbital boost momentum $j$,
generating boost transformations for the zero-modes, is given by
\be
\label{jtw}
j  \equiv i (x^+ \pa_+ - x^- \pa_-) = (M^2-\tM^2)/(2\nu)
\ee
and fixed by the level matching condition $L_0-\bar L_0=0$.
It is furthermore quantized in integer multiples of $1/\beta$, in order for
the twisted state to be invariant under the orbifold projection.
It will be useful to further define the two-dimensional square
mass $\mu^2=(M^2+\tM^2)/2$, which, unlike $j$, takes continuous values.

The coordinate $\xp$ introduced in \eqref{reals} describes
the location  of the winding string in the Heisenberg
representation (i.e. at $\tau=0$), since the zero-mode part
of $X^\pm(\sigma,\tau)$ may be written as
\be
\label{x0}
 X^{\pm}_0(\sigma,\tau) =
e^{\mp \nu \sigma} \left[ \cosh (\nu \tau)
~  x^\pm \pm \frac{2i}{\nu} \sinh (\nu \tau) ~
\pa_\mp \right]
\ee
Notice that the dependence of the zero-mode on the $\sigma$
coordinate amounts to a boost in the $(x^+,x^-)$ plane,
i.e. a translation along the compact coordinate of the
orbifold.

Alternatively, it is possible to use a representation
where half of the covariant derivative operators
are diagonalized, e.g.
\be \a^-=i\nu\partial_{\a^+},\ \ \
\talpha^+=i\nu\partial_{\talpha^-} \label{oscrin}
\ee
acting on functions of the variables $\a^+,\ta^-$ taking values
in the quadrant $\Real^{\eps}\times \Real^{\teps}$.
On-shell wave  functions are now powers of their arguments,
\be \label{eigoin} f( \a^+,  \talpha^- ) = N_{in}
\left( \eps ~ \a^+ \right)^{\frac{M^2}{2i\nu} -
\frac12} \left( \teps~ \ta^-
\right)^{\frac{\tM^2}{2i\nu} - \frac12} \ee
The notation $N_{in}$ for the normalization factor,
to be computed below, anticipates the fact that this representation
is appropriate to describe an {\it in} state.
This representation is the analogue of the oscillator
representation for the standard harmonic oscillator, however,
since the powers are not integers, it involves four different
branches, according to the signs $\eps$ and $\teps$ of
$\a^-$ and $\ta^+$. These gives rise to qualitatively distinct classes
of twisted sector states:
\begin{itemize}
\item {\it  Short strings}: for $\eps\teps=1$ the strings wind around the
Milne spatial circle, and propagate from infinite past to infinite
future. When $j\neq 0$, they also extend in the Rindler regions to a finite
distance $r_-^2=(M-\tilde M)^2/(4\nu^2)$.
\item {\it Long strings}: for $\eps\teps=-1$, the strings 
instead live in the Rindler regions only,
and  correspond to static configurations which extend from spatial
infinity in L or R to a finite distance $r_+^2=(M+\tilde M)^2/(4\nu^2)$,
and folding back to infinity again. In addition, there exists
tunneling configurations whereby a long string dissolves
at a distance $r_+$ and  reemerges as a short  string at distance
$r_-$, before escaping into the cosmological regions (P and F).
\end{itemize}
In particular, in contrast to Euclidean orbifolds, twisted strings
are in no sense localized near the singularity.

The oscillator representation \eqref{oscrin} can be related
to the real-space representation \eqref{reals} via
\be f(x^+,x^-) =
\int d\talpha^+ d\alpha^- \Phi^{in}_{ \nu, \ta^+,
\alpha^-}(x^+,x^-) f( \a^+, \ta^-) \ee
where the kernel is given  by
\be \label{phiin} \Phi^{in}_{ \nu,
\a^+,
\talpha^-}(x^+,x^-) = \exp\left( \frac{i\nu x^+
x^-}{2} - i \a^+ x^- -i \talpha^- x^+ +
\frac{i}{\nu} \a^+ \talpha^- \right) \ee
This kernel may be viewed as the wave function for an off-shell
winding state with momenta $\a^+$  and $\ta^-$.
Equivalently, one may diagonalize the complementary set of operators,
\be \a^+=-i\nu\partial_{\a^-},\ \ \
\talpha^-=-i\nu\partial_{\talpha^+} \label{oscrout}
\ee
leading to on-shell wave  functions
\be \label{eigoout} f( \a^-,  \talpha^+ ) = N_{out}
\left( \eps ~ \a^- \right)^{-\frac{M^2}{2i\nu} -
\frac12} \left( \teps~ \ta^+
\right)^{- \frac{ \tM^2}{2i\nu} - \frac12} \ee
Those are related  to the real-space representation  by the
kernel
\be \label{phiout} \Phi^{out}_{ \nu, \talpha^+,
\alpha^-}(x^+,x^-) = \exp\left( - \frac{i\nu x^+
x^-}{2} - i \ta^+ x^- -i \a^- x^+ -
\frac{i}{\nu} \ta^+ \a^- \right)
\ee

\subsection{Two-point function of twisted fields}
As usual, the two-point function of twisted fields vanishes unless
the two fields carry the same winding number. In addition, since
the excited modes are quantized as in flat space, the amplitude
is non-zero only when the excitation levels on the left and
on the right-moving side are equal. Throughout this section we will
limit ourselves to the case of twisted strings without internal
excitation, but with some internal momentum $k_i$ so that the
resulting state is non-tachyonic in 2 dimensions. For generality
however, we will not use any specific value for the two-dimensional
masses $M^2$ and $\tM^2$.

We will find it convenient to use the representations
\eqref{eigoin} and \eqref{eigoout} (or rather, its hermitian conjugate)
for  the  {\it in}  and {\it out} twisted states, respectively.
The inner product between {\it in} and {\it out} states
may be simply determined by returning to the
real-space representation and using \eqref{hermi}:
\be
_{out}\langle f_1 | f_2 \rangle_{in} =
\int d\talpha^+ d\alpha^- d\alpha^+ d\tilde \alpha^-
f_1^*(\tilde \alpha^+, \alpha^-) ~
e^{\frac{i}{\nu}\left(\alpha^+ \alpha^- +  \tilde\alpha^+
    \tilde\alpha^-  \right)}
f_2(\alpha^+, \tilde \alpha^-)
\label{2pt}
\ee
where the integration domain is $
\Real^{\eps_1}\times\Real^{\teps_1}\times\Real^{\eps_2}\times\Real^{\teps_2}$.
The overlap may be computed straightforwardly in terms of Gamma
functions, and yields
\bea
_{out}\langle f_1 | f_2 \rangle_{in} &=& N_{out}^* N_{in}
\left[ 2i\nu^2~
(i\nu\eps_1\eps_2)^{\frac{M_2^2}{2i\nu} - \frac12}
\Gamma\left( \frac{M_2^2}{2i\nu} + \frac12 \right)
\delta(M_1^2-M_2^2) \right] \nn\\
&&\times
\left[ 2i\nu^2~
(i\nu\teps_1\teps_2)^{ \frac{\tM_1^2}{2i\nu} - \frac12}
\Gamma\left( \frac{\tM_1^2}{2i\nu} + \frac12 \right)
\delta(\tM_1^2-\tM_2^2) \right]
\label{overlap}
\eea
where the first bracket corresponds to the $(\a^+,\a^-)$ integral
and the second to the $(\ta^+,\ta^-)$ integral.
Due to the orbifold projection however,
the momentum $j=(M^2-\tM^2)/(2\nu)$ is a discrete quantity, hence
its conservation should be enforced by a Kronecker symbol rather
than Dirac's delta function. The way to remedy this problem is
to mod out the integration domain on $(\a^\pm,\ta^\pm)$ by the same boost
which identified the $x^\pm$ coordinates,
\be
\label{boosta}
(\a^\pm,\ta^\pm) \equiv (e^{\pm 2\pi\beta} \a^\pm, e^{\pm 2\pi\beta}  \ta^\pm)
\ee
This action clearly leaves the on-shell wave functions with quantized $j$
\eqref{eigoin},\eqref{eigoout} invariant, and is consistent with the kernels
\eqref{phiin},\eqref{phiout}\footnote{It may seem that one should mod out
the $\a^\pm$ and $\ta^\pm$ planes by independent boosts, however this
would lead to a conservation of both $M^2$ and $\tM^2$ by Kronecker symbols,
inconsistent with the continuous nature of $\mu^2$.}. Under this prescription,
and upon choosing the normalization factors
\be
N_{in}= N_{out}^* =
(2\nu)^{-1/2}~ (i\nu)^{ - \frac{M^2+\tM^2}{4i\nu} - \frac12}
\left[\Gamma\left( \frac{M^2}{2i\nu} + \frac12 \right)
\Gamma\left( \frac{\tM^2}{2i\nu} + \frac12 \right) \right]^{-1/2}
\ee
we obtained delta-normalized states,
\be
_{out}\langle f_1 | f_2 \rangle_{in}
=\delta(\mu_1^2-\mu_2^2) ~\delta_{j_2-j_1} (\eps_1\eps_2)^{\frac{M_2^2}{2i\nu} - \frac12}
(\teps_1\teps_2)^{\frac{\tM_1^2}{2i\nu} - \frac12}
\ee
For simplicity, we will consider only amplitudes involving the same type of twisted states
for the $in$ and $out$ states $\eps_1=\eps_2$ and $\teps_1=\teps_2$.

Finally, for the purpose of computing more general overlaps, it will
be useful to use generating function techniques, and view
the on-shell {\it in} wave function \eqref{eigoin} as a derivative
w.r.t. generating parameters $(\zeta^-,\tzeta^+)$ {\it with continuous order},
\be
\label{derin}
f(\a^+,\ta^-) =
{\cal D}^{in}_{M^2,\eps,\teps}\cdot
e^{-i \a^+  \zeta^- - i \ta^- \tzeta^+}  \vert_{\zeta^-=\tzeta^+=0}
\ee
where
\be
{\cal D}^{in}_{M^2,\eps,\teps} \equiv N_{in}~
\left[ i \eps \frac{\pa}{\pa \zeta^-} \right]^{\frac{M^2}{2i\nu}-\frac12}
\left[ i \teps \frac{\pa}{\pa \tzeta^+} \right]^{\frac{\tM^2}{2i\nu}-\frac12}
\ee
In this expression, the continuous derivative operator
is defined by\footnote{E.g,
acting on $f(x)=e^{ax}$ yields $a^\lambda$ as it should. }
\be \left.\frac{\partial^{\lambda}}{\partial x^{\lambda}}
f(x) \right\rvert_{x=0} =\frac{(-1)^\lambda}{\Gamma(-\lambda)}
\int_0^\infty dx\, x^{-\lambda-1} f(x)  \ ,\label{difdef}
\ee
Similarly, the {\it out} wave function may be represented as
a derivative with respect to a second  set of parameters $(\zeta^+,
\tzeta^-)$,
\be
\label{derout}
f(\ta^+,\a^-) =  {\cal D}^{out}_{M^2,\eps,\teps} \cdot
e^{-i \ta^+  \tzeta^- - i \a^- \zeta^+}\vert_{\zeta^+=\tzeta^-=0}
\ee
where
\be
{\cal D}^{out}_{M^2,\eps,\teps} \equiv N_{out}
\left[ i \teps  \frac{\pa}{\pa \tzeta^-} \right]^{-\frac{\tM^2}{2i\nu}-\frac12}
\left[ i \eps \frac{\pa}{\pa \zeta^+} \right]   ^{-\frac{M^2}{2i\nu}-\frac12}
\ee
For simplicity, we will drop the indices from the operators ${\cal D}$
when no confusion is  possible. The
integral \eqref{2pt} over $\a^\pm, \ta^\pm$ is now Gaussian\footnote{The
peculiar integration domain does not make any difference for this argument.},
\be
\label{2ptz}
\langle f_1 | f_2 \rangle = \nu^2 ~
{\cal D}_1^* {\cal D}_2 \cdot
e^{i \nu (\zeta^+ \zeta^-+\tzeta^+ \tzeta^-)
}
\vert_{\zeta^\pm=\tzeta^\pm=0}
\ee
and the overlap \eqref{overlap} follows from the formal derivation rule
\be
\left[ i \eps_1 \frac{\pa}{\pa \zeta^+}\right]^{\frac{M_1^2}{2i\nu}-\frac12}
\left[ i \eps_2 \frac{\pa}{\pa \zeta^-}\right]^{\frac{M_2^2}{2i\nu}-\frac12}
e^{ i \nu \zeta^+ \zeta^- }\vert_{\zeta^\pm=0}
=2 i\nu^2  (i\nu \eps_1 \eps_2)^{\frac{M_2^2}{2i\nu} - \frac12}
\Gamma\left( \frac{M_2^2}{2i\nu} + \frac12 \right)
~\delta(M_1^2-M_2^2)
\ee
which generalizes to continuous order the familiar fact for
ordinary (integer order) derivatives,
\be
\left[ i \frac{\pa}{\pa \zeta^+}\right]^{n}
\left[ i \frac{\pa}{\pa \zeta^-}\right]^{m}
e^{ i \nu \zeta^+ \zeta^- }\vert_{\zeta^\pm=0}
=(i\nu)^{n} ~\Gamma(n+1) ~\delta_{m-n}
\ee
Notice that the ``generating'' $(\zeta^+,\tzeta^-)$ representation is simply
obtained by Fourier transform from the  $(\alpha^-,\talpha^+)$
oscillator representation, hence is equivalent to the  $(\alpha^+,\talpha^-)$,
up to a rescaling by $\nu$.

\subsection{Untwisted vertex operators and stringy fuzziness}
Tree-level scattering amplitudes involving two twisted
states and an arbitrary number of untwisted states can be computed
by operator methods, by choosing a Lorentzian world-sheet which is an
infinite cylinder, in the ${\it in}$ and ${\it out}$ twisted vacua
at $\tau=\pm\infty$. For this, it is important to evaluate the action
of untwisted vertex operators on the twisted Hilbert
space.

\subsubsection{Tachyon vertex operator}
The tachyon vertex operator may be defined
in the untwisted sector by the normal ordered expression
\be
\label{vt}
V_T =
:e^{i k^+ X^- + k^- X^+}:(z)
\equiv \lim_{w\to z}
e^{i k^+ X^-}(w)~
e^{i k^- X^+}(z) ~
e^{k^+ k^- [X_{>0}^-(z), X^+_{<0}(w)]}
\ee
where $X^{\pm}_{\gtrless}$ denotes the positive and negative frequency
parts of $X^{\pm}$, as defined by the untwisted mode expansion.
Notice that the operators $e^{i k^\pm X^\mp}$ do not need any normal ordering
prescription as there are no short distance singularities between
$X^\pm$ and itself. In order to obtain a state invariant under
the orbifold action, it is necessary to further smear
\eqref{vt} over the action of the
boost $k^\pm \to e^{\pm v} k^\pm$, but we will refrain from
explicitly doing so in order to avoid cluttering. The definition above is the same as the familiar CFT normal ordering definition
\be
V_T\equiv \lim_{w\to z}~ e^{i k^+ X^-}(w)~
e^{i k^- X^+}(z) ~ e^{k^+k^-log(w-z)}
\ee
which we can then easily transfer to other sectors.

In terms of the twisted mode expansion, the normal ordered (with respect to the twisted sector vacuum)
tachyon vertex operator $V_T$ now reads
 \be
V_T =
e^{i ( k^+ X^-_{\prec 0} + k^- X^+_{\prec 0})}~
e^{i ( k^+ X^-_{\succ 0} + k^- X^+_{\succ 0})}~
e^{i ( k^+ X^-_{0} + k^- X^+_{0})}~
e^{k^+ k^- \left(
[X_{>0}^-, X^+_{<0}]
-[X_{\succ 0}^-, X^+_{\prec 0}] \right) }
\ee
where we have separated the the $X$'s into positive and negative
frequency parts $X^{\pm}_{\succ,\prec}$ of $X^{\pm}$ as defined by the twisted mode
expansion. $X_0^\pm$ denotes the twisted quasi-zero-mode as in
\eqref{x0}. The difference of commutators is now finite as
$w\to z$, and evaluates to
\be
\Delta(\nu)\equiv
[X_{\succ 0}^-, X^+_{\prec 0}]- [X_{>0}^-, X^+_{<0}] =
\sum_{n>0} \left(  \frac{2}{n} - \frac{1}{n+i\nu}
- \frac{1}{n-i\nu} \right)
\ee
The infinite sum can be evaluated in terms of the digamma function
$\psi(x)=d\log\Gamma(x)/dx$,
\be
\Delta(\nu) = \psi(1+i\nu)+\psi(1-i\nu)-2\psi(1)
\ee
In particular, this implies that
the excited states contribute a non-trivial form factor\footnote{Form factors
in usual $Z_N$ Euclidean orbifolds have been discussed in
Ref. \cite{GrootNibbelink:2003zj}} for
untwisted strings in the background of a twisted string:
\be
\label{exctac}
\langle \nu | :e^{i (k^+ X_{ex}^- + k^- X_{ex}^+)}:  | \nu \rangle
= \exp\left( - k^+ k^- \Delta(\nu) \right)
\ee
Notice that this result is unaffected by the
projection on boost invariant states. We will return to the important
implications of this form factor in Section 2.5 below, after discussing
the graviton vertex operator.

\subsubsection{Gauge boson and graviton vertex operators}
We now turn to the gauge boson vertex operator; while this state
does not satisfy the closed string level matching condition,
it is a useful warm-up before tackling the graviton.

In the untwisted sector, the gauge boson vertex operator may be
defined in terms of the tachyon vertex, as\footnote{Recall that operators
are implicitly radially ordered.}
\be
\label{va}
V^{\mu}_A =
:z \partial_z X^\mu e^{ikX}:~
\equiv \lim_{w\to z}
\frac12 \left[ z \partial_z X^\mu(z) ~V_T(w)
+ V_T(z) ~w \partial_w X^\mu(w) \right]
\ee
This differs from the more standard definition
\be
V^{'\mu}_A
\equiv \lim_{w\to z}
\frac12 \left[ z \partial_z X^\mu(z) ~V_T(w)
-\frac{i\,z\,k^\pm}{z-w} \right]
\ee
by a total derivative term proportional to $k^\pm~\partial_z V_T$,
and leads to the same results when
restricting to physical  polarizations $\zeta_\mu$  such that
$\zeta_\mu k^\mu=0$.
Nevertheless, the
symmetric ordering prescription \eqref{va} has
the advantage  that the vertex operator for a
longitudinal gauge boson becomes a total derivative,\footnote{Checking
this relation requires a  careful handling of the zero-modes.}
\be
i k_\mu V^{\mu}_A = z \partial_z :e^{ikX}:
\ee
implying current conservation. While the non-conserved part of an
electromagnetic current
current does not source photons in the Lorentz gauge $\pa_\mu
A^\mu=0$, we find it more convenient to maintain current conservation
throughout.

Expressing $V_A$ in terms of the twisted
mode expansion, we find the normal ordered expression
\be
V^{\pm}_A =
-i \sum_{n>0}  \left[ \alpha^\pm_{-n} ~V_T ~z^{n}
+ V_T ~\alpha^\pm_{n} ~z^{-n} \right] z^{\mp i\nu}
-\frac{i}{2} \left[ \alpha^\pm_{0} ~V_T
+ V_T ~ \alpha^\pm_{0}  \right] z^{\mp i\nu}
\mp \nu {k^\pm} ~V_T
\ee
where the last term comes from normal ordering the excited modes of
$\pa_z X^\pm$ through $V_T$. Notice that it amounts to shifting
$\alpha_0^\pm \to \a_0^\pm \pm i \nu k^\pm z^{\pm i \nu}$.
Similarly, computing the anti-holomorphic derivative
one finds a shifting $\talpha_0^\pm
\to \ta_0^\pm \mp i \nu k^\pm z^{\pm i \nu}$.
Importantly, the zero-mode part of the vertex operator appears to be
symmetrized, as a requirement from current conservation.

Having dealt with the gauge boson, we can now obtain the vertex
operator for the graviton, dilaton and Kalb-Ramond two-form by
applying the same procedure on the holomorphic and anti-holomorphic
sides:
\bea
V^{\mu\nu}_G &=&
:z \pa_z X^\mu {\bar z} ~\pa_{\bar z} ~X^\nu ~e^{ikX}:~ \\
&\equiv& \lim_{w\to z,\bar w\to \bar z}
\frac14 \left[
z \pa_z X^\mu ~{\bar z} \pa_{\bar z} X^\nu ~V_T(w,\bar w)
+~V_T(z,\bar z) w \pa_w X^\mu(w) ~{\bar w} \pa_{\bar w} X^\nu \right.\nn\\
&&\left.
+z \pa_z X^\mu ~V_T(w,\bar z)~{\bar w} \pa_{\bar w} X^\nu
+{\bar z} \pa_{\bar z} X^\nu  ~V_T(z,\bar w)~w \pa_w X^\mu
\right]
\eea
Notice in the above expression that we radially ordered
independently on the holomorphic and anti-holomorphic sides,
and did not symmetrize in the polarization indices $(\mu,\nu)$:
the dilaton, graviton and Kalb-Ramond vertex operators can be obtained
by extracting the trace from the ten-dimensional polarization
tensor $\zeta_{\mu\nu}$ and (anti)symmetrizing.
Inserting the twisted mode expansion, we obtain
a normal ordered expression with respect
to the positive/negative frequency excited twisted modes, while
the quasi-zero-modes are symmetrized and shifted according to
\be
\label{sha}
\alpha_0^\pm \to \a_0^\pm \pm i \nu ~k^\pm~ z^{\pm i \nu}\ ,\quad
\talpha_0^\pm
\to \ta_0^\pm \mp i \nu~ k^\pm~ z^{\pm i \nu}
\ee
The energy-momentum tensor is then by construction conserved.

\subsection{Stringy fuzziness with a twist \label{fuzz}}

The form factor \eqref{exctac}  appears in fact generically for
all untwisted vertex operators, as a result of normal ordering
in the twisted vacuum, and has important
physical consequences, as we now explain.

It is of course well-known that, due to zero-point fluctuations,
off-shell strings have a logarithmically diverging r.m.s.
size \cite{Karliner:1988hd}:
for a scattering process of energy $E$, strings grow as large as
$\sqrt{\log E}$, implying the standard Regge behavior $A\sim s^{t} \sim
e^{t\log s}$. These divergences are usually absorbed by non-local
wave function renormalization (i.e. normal ordering of the vertex operators),
leading to local $n$-point functions at low energy.

In contrast, in the Lorentzian orbifold case, we find that zero-point
fluctuations have a non-trivial dependence on the winding sector $w$,
and therefore cannot be reabsorbed in a field redefinition of the
untwisted states (nor can it be absorbed into a simple redefinition of the
twisted states). Instead, as apparent from \eqref{exctac}, the
presence of a winding string polarizes\footnote{The same polarization
effect takes place for charged open strings in an electric field,
due to the analogy noted in \cite{Pioline:2003bs}.}
untwisted string states into a cloud of r.m.s. size $\sqrt{\Delta(\nu)}$.
This size is proportional to that of the winding string
at small $\nu$, at large winding
number however it grows logarithmically with $\nu$:
\bea
\Delta(\nu)&=&2\zeta(3) \nu^2 + O(\nu^4)\\
& =& 2\log \nu -\frac{23}{20} + O(\nu^{-2}).
\eea
Importantly, in contrast to usual finite order Euclidean rotation
orbifolds, the winding number is unbounded, and can lead to
a fuzziness much larger than the string scale.

While the complete form factor depends on the zero-mode contributions
as well as the excitation levels, it is instructive to consider the effect
of the universal exponential factor \eqref{exctac} in real space.
Since \eqref{exctac} diverges exponentially in the space-like region
$k^+ k^-<0$, it is necessary to assume that the remaining contribution
$A_0(k^+,k^-)$ of zero-modes and oscillators is sufficiently suppressed
in this region. If so, the effect of \eqref{exctac} is to convolute
$A_0(x^+,x^-)$ with the Fourier transform of \eqref{exctac},
\be
A(y) = \int d^2 y ~
\exp\left( - \frac{(y^+-x^+)(y^- -x^-)}{\Delta(\nu)} \right) A_0(x)
\ee
or, equivalently, to apply the non-local differential operator
\be\label{operator}
A(x) = \exp\left( \Delta(\nu) \pa_+ \pa_- \right) ~ A_0(x)
\ee
This non-locality is reminiscent of non-commutative star products,
but also markedly different, as it affects the amplitude rather than
the phase of the wave function, and is already non-trivial at the
level of single particle states. The analogy to the charged open
string\cite{Pioline:2003bs}  makes it clear that this non-locality is
to charged particles in an electric field
what the non-commutative Moyal product
is to neutral dipoles in a magnetic field. The important point however
is that the ``non-commutativity'' arises here in the closed string sector,
in the vicinity of a cosmological singularity. It would be interesting
to develop a semi-classical picture of this non-locality,
in analogy with the rigid dipole picture for the standard
non-commutativity in a magnetic field \cite{Bigatti:1999iz}.

The simplest example is a Gaussian
distribution peaked at $(x_0^+,x_0^-)$ with r.m.s. size $a$,
\be
\label{test1}
A_0(x)=\exp\left(-\frac{(x^+-x_0^+)^2+(x^--x_0^-)^2}{a^2}
\right)
\ee
whose transform $A(x)$ is
\be
\frac{a^2}{\sqrt{a^4-4\Delta^2}}
\exp\left(-\frac{a^2}{a^4-4\Delta^2}
\left[
(\xp-x_0^+)^2+(\xm-x_0^-)^2+\frac{4\Delta}{a^2}(\xp-x_0^+)(\xm-x_0^-)
\right]\right)\ ,
\ee
The latter is still localized around $x_0^\pm$, but deformed
into an ellipse. The eigenvalues of the quadratic form are
$1/(a^2-2\Delta)$ and $1/(a^2+2\Delta)$ and the main effect is to
squeeze the r.m.s. fluctuations in one direction to zero as
$\Delta$ approaches $a^2/2$ (the direction that
shrinks to zero eventually is parallel to $(x^+,x^-)\propto
(1,-1)$ for $\Delta>0$).

Another example, more relevant for boost invariant situations,
is the test function
\be
\label{test2} A_0(x^+,x^-) = \exp\left( - \frac{(x^+ x^- +
r_0^2)^2}{a^2\,r_0^2} \right)
\ee
which, for real $r_0$,
is localized around a finite distance $r_0$ in the Rindler
wedges, with r.m.s size $a$. This can viewed as a slice of a ``long
string", localized around $r_0$ and winding around the compact
Rindler time (for $r_0^2<0$, it instead describes a short string
winding the Milne circle at a fixed time $i r_0$).

The action of the operator \eqref{operator} on this test function can not be computed in a closed form, but we can extract the main features
by studying the asymptotic behavior using steepest descent
methods. First we diagonalize one of the derivative operators, say $\pa_+$,
by Fourier transform in $x^+$,
\bea
A(x)&=&\frac1{2\pi}\int_{-\infty}^\infty dp^-\,e^{-i\Delta p^-\pa_-}\int _{-\infty}^\infty dy^+\,e^{+ip^-(y^+-x^+)}\exp\left(-\frac{\left(y^+x^-+r_0^2\right)^2}{a^2r_0^2}\right)\\
    &=&\frac{a\,r_0\sqrt{\pi}}{2\pi}\int_{-\infty}^\infty dp^-\,e^{-i\Delta p^-\pa_-}\frac{e^{-ip^-x^+}}{\abs{x^-}}
    \exp\left[-p^-\left(\frac{r_0}{2x^-}\right)^2\left(a^2p^-+4i x^-\right)\right]
\eea
The action of $~e^{-i\Delta p^-\pa_{x^-}}~$ is to shift the
$x^-$ coordinate by an amount $~-i\Delta p^-~$. After
changing the integration variable to $q=\Delta p^-/ \xm$, we find
\begin{equation}\label{test2int}
A(x)    =\frac{a\,r_0\sqrt{\pi}}{2\pi\Delta}\int _{-\infty}^\infty dq\,\frac{e^{-i\frac{\xp\xm}\Delta q}}{\abs{1-iq}}
    \exp\left[-\frac{r_0^2}{4\Delta}\frac{q}{(1-iq)^2}\left(\frac{a^2}{\Delta}q+4i(1-iq)\right)\right]
\end{equation}
The integral above is well defined as can be seen by
deforming the integration contour into the lower half
$q$-plane for $x^+x^->0$, i.e. for an observer in the future Milne region. The
denominator is written as $\abs{1-iq}\rightarrow\sqrt{1+q^2}$, paying the cost
of a branch cut on the imaginary axis extending from $-i$ to $-i\infty$,
enclosed by the integration contour around
(the other branch cut stretching from $+i$
stretching up to $+i\infty$ is not relevant for this region).
The integral is dominated by three saddle points whose
qualitative behavior changes between small\footnote{Recall, however,
that for a fixed boost parameter we can take $\Delta$ to infinity but not
to zero, since $\nu$ is bounded from below}  and large
$\Delta/a^2$.
We will explore these regions separately.

For large $\Delta/a^2$ (i.e. large stringy fuzziness)
the saddle points are (to leading order in $a^2/\Delta$):
\begin{equation}
    q_1=-i+i\frac{a^2}{2\Delta}\hspace{1cm},\hspace{1cm}
    q_{2,3}=-i\mp\sqrt{\frac{r_0^2}{\xp\xm}}
\end{equation}
Using these saddle points we can estimate \eqref{test2int} in this limit:
\begin{equation}
   A(x) =\sim \frac{a^3\,e^{-\frac{r_0^2}{a^2}-\frac{\xp\xm}{\Delta}}}{4\Delta^{\frac32}}
    +\frac{a\,r_0\,e^{\frac{r_0^2\xp\xm}{\Delta}}}
{\sqrt\Delta\left[\xp\xm(r_0^2+\xp\xm)\right]^{\frac14}}\,\cos\left(\frac\Theta4+2\frac{r_0\sqrt{\xp\xm}}{\Delta}\right)
\end{equation}
where $\Theta=\tan^{-1}\left(2\sqrt{\xp\xm/r_0^2}\right)$.
The saddle points are reliable in the limit of large
$r_0^2 x^+x^-/\Delta^2$. Furthermore, the main contribution is the second
(coming from $q_{2}\,,\,q_3$). The smeared
function is therefore concentrated in a
region of the past/future wedges around $r_0$ with r.m.s. size
$\sqrt\Delta$, much larger than the original size of order $a$.

For small $\Delta/a^2$ instead (i.e. in the semi-classical limit), the saddle
points are located, to leading order in $\Delta/a^2$, at
\begin{equation}
    q_1=-2i\frac\Delta{a^2}\left(1+\frac{\xp\xm}{r_0^2}\right)\hspace{1cm},\hspace{1cm}
    q_{2,3}=-\frac32\,i
\mp\left(\frac{a^2}{\Delta}\,\frac{r_0^2}{2\xp\xm}\right)^{1/2}
\end{equation}
giving a saddle point estimate
\begin{equation}
   A(x)\sim e^{-\frac{(\xp\xm+r_0^2)^2}{a^2 r_0^2}}+
    \left(\frac{2a^2r_0^2}{\Delta\xp\xm}\right)^{\frac14}
e^{\frac{a^2r_0^2}{4\Delta^2}}\cos\left(\frac\pi4+
\frac{a~r_0\,\sqrt{2x^+x^-}}{\Delta^{\frac32}}\right)
\end{equation}
One recovers in the first term the original test function $A_0(x)$,
from the contribution of the saddle point $q_1$. However, the second
term, which comes from the saddle points $q_2,q_3$, overwhelms the
first in the limit $\Delta\to 0$. Although the saddle points $q_{2,3}$
move off to infinity in this limit, we have not found any indication
that they should not contribute to the contour integral. If correct,
this implies that the limit $\beta\to 0$ is singular, against expectations
from usual decoupling arguments. Strictly speaking, however,
in order to obtain a finite string background we need to take
$x^+ x^-\to \infty$ at the same time as $\beta\to 0$, holding $x^+x^-\beta^2$
fixed. It would be interesting to investigate this limit in detail.

\section{Scattering amplitudes with two twisted strings}
We now turn to tree-level scattering amplitudes involving two twisted
strings, and one or more untwisted states. For this purpose, it is
convenient to view  the string world-sheet as an infinitely long
cylinder , with the twisted vertex operators being inserted at
$\tau=\pm\infty$. Amplitudes can then simply be computed by
operator methods, in the Hilbert space of the twisted string.
In this section, we compute the scattering amplitude of
of one tachyon or graviton off a short or a  long
winding string. We also study the four point function of
two twisted and two untwisted strings, and find that it diverges
as a result of large winding string emission.

\subsection{Two-twisted one-untwisted amplitude}
We now turn to the amplitude involving two twisted strings of
(identical) winding number $w$, and one untwisted string.
We will start by choosing the latter as a tachyon, although
our interest eventually lies in the case of a graviton.
As before, the twisted strings are assumed to be in their ground
state, and on shell. We do not impose the mass-shell
condition for the untwisted state at this stage, however.

\subsubsection{Tachyon form factor}

Since the twisted strings are assumed to be on-shell, the amplitude
is independent of the location of the untwisted vertex operator,
which we choose to be at $\tau=\sigma=0$ (i.e. $z=\bar z=1$).
The amplitude is thus given by
\be
\label{123}
\langle 1 | T(2) | 3 \rangle
=\int_{-\infty}^{\infty}
dv_2 ~\langle 1 | :e^{i (p_2^+(v_2) X^- + p_2^-(v_2) X^+  +  p_2^i X^i) }: | 3
\rangle~e^{i j_2 v_2}
\ee
where $p_2^\pm(v) =  e^{\mp v}  p_2^\pm$. The integral over $v_2$
projects the tachyon vertex operator to a boost-invariant state
with integer momentum $j_2$.
It is easy to see that the integral
over $v_2$ contributes a momentum-conserving
Kronecker symbol\footnote{Switching the boost identification
\eqref{boosta} from the oscillator variables to $v$ effectively reduces
the range of $v$ to the interval $[0,2 \pi \beta]$, leading to
a Kronecker symbol rather than a Dirac delta.}
$\delta_{-j_1+j_2+j_3}$, where the boost momentum
of  the  twisted states $1$  and $3$ is given by \eqref{jtw}.
We will therefore set $v_2=0$ for  now and reinstate momentum
conservation at the end
of the computation.  Similarly,  the transverse part  yields
a delta function enforcing transverse momentum conservation,
$\delta(-\vec p_1+\vec p_2+\vec p_3)$. The amplitude now factorizes
as a product of a zero-mode and excited-mode piece,
\be
\langle 1 | e^{i p_2^+ X^- + i p_2^- X^+ } | 3
=
_{out}\langle f_1 | e^{i p_2^+ X_0^- + i p_2^- X_0^+}  | f_3 \rangle_{in}
\times
\langle \nu | e^{i p_2^+ X_{ex}^- + i p_2^- X_{ex}^+}  | \nu \rangle
\ee
where the  contribution of the excited
modes was already computed in \eqref{exctac}.

We now turn to the zero-mode part. Using the real space representation
\eqref{reals}, the zero-mode contribution reduces to the overlap
\be
\label{overc}
_{out}\langle f_1 | e^{i (p_2^+ X_0^- + p_2^- X_0^+)}  | f_3
\rangle_{in}
=\int dx^+ dx^- f_1^*(x^+,x^-) e^{i (p_2^+ x^- + p_2^- x^+)}
f_3(x^+,x^-)
\ee
where $f_1$ and $f_3$ are eigenmodes of the Klein-Gordon equation
with electric charge $\nu$. Were it not for the contribution of the
excited modes \eqref{exctac}, the tachyon form factor evaluated
at a point $(x^+,x^-)$ would therefore be equal to the product
of the twisted wave functions $f_1^*(x^+,x^-) f_3(x^+,x^-)$ at that point.
Instead, as described in Section \ref{fuzz}, the tachyon form
factor will be smeared over a size $\sqrt{\Delta}$.

In order to evaluate the overlap \eqref{overc} for on-shell
twisted wave functions, let us use the
{\it in} and {\it out} oscillator representations for $f_3$ and $f_1$,
respectively.
Replacing $X_0^\pm$ by its expression \eqref{x0} in terms of the quasi
zero-modes, the amplitude factorizes into a product of
expectation values in the left and right-moving zero-mode sectors,
\be
_{out}\langle f_1 | e^{i (p_2^+ X_0^- + p_2^- X_0^+)}  | f_3 \rangle_{in}
 = e^{\frac{p_2^+ p_2^-}{i\nu}}
\langle M_1 \rvert
e^{-\frac{i}{\nu} p_2^+ \alpha^-}
e^{ \frac{i}{\nu} p_2^- \alpha^+}
\lvert M_3 \rangle ~
\langle \tM_1 \rvert
e^{-\frac{i}{\nu} p_2^- \ta^+ }
e^{+\frac{i}{\nu} p_2^+ \ta^-}
\lvert \tM_3 \rangle
\ee
where we used the Baker-Campbell-Hausdorff
formula 
to split each exponential into a product of diagonalized operators.
Recalling that the inner product in the oscillator representation
is given by \eqref{2pt}, we obtain
\begin{multline}
\langle f_1 | e^{i (p_2^+ X_0^- + p_2^- X_0^+)}  | f_3 \rangle
= e^{\frac{p_2^+ p_2^-}{i\nu}}~
\int d^4\alpha~
f_1^*(\tilde \alpha^+, \alpha^-) \\
e^{\frac{i}{\nu}\left[\alpha^+ \alpha^- +  \tilde\alpha^+ \tilde\alpha^-
+ p_2^- (\a^+ - \ta^+)
- p_2^+ (\a^- - \ta^-)\right]}
f_3(\alpha^+, \tilde \alpha^-)
\label{3pt}
\end{multline}
Representing the {\it in} and {\it out} wave functions as continuous
derivatives with respect to generating parameters as in \eqref{derin},
\eqref{derout}, the integral over $\a^\pm,\ta^\pm$ is Gaussian,
leading to
\be
\label{3ptz}
\langle f_1 | e^{i (p_2^+ X_0^- + p_2^- X_0^+)}  | f_3 \rangle
=
\nu^2 ~{\cal D}_1^* ~{\cal D}_3 \cdot
e^{i \nu (\zeta^+ \zeta^-+\tzeta^+ \tzeta^-)
- i (\zeta^+ + \tzeta^+) p_2^-
-i (\zeta^- + \tzeta^-) p_2^+
+\frac{i}{\nu}{p_2^+ p_2^-}
}
\vert_{\zeta^\pm=\tzeta^\pm=0}
\ee
We may now evaluate the continuous derivatives by using the
identity\footnote{This identity follows from
using the continuous derivative operator \eqref{difdef}, or equivalently
performing the $\alpha^\pm$ integral explicitly, and using
the transformation property
$U(\lambda,\mu,z)=z^{1-\mu}U(1+\lambda-\mu,2-\mu,z)$.
In particular, even
when $M_1^2/(2i\nu)-1/2=n$ and $M_2^2/(2i\nu)-1/2=\bar n$ are integer,
it does {\it not} reduce to the analytic continuation of the
Euclidean result, which would instead be given by an hypergeometric
function $U(\mu-\lambda,\mu;-ip^+p^-/\nu)$, or equivalently by the
generalized Laguerre polynomial $L_{\bar n}^{n-\bar n}(-ip^+p^-/\nu)$.
Notice that the parameter $\mu$ in \eqref{Udef} should not be confused
with the two-dimensional squared mass $\mu^2$.}
\begin{multline}
\label{idw}
\left[ i \frac{\pa}{\pa \zeta^+}\right]^{\frac{M_1^2}{2i\nu}-\frac12}~
\left[ i \frac{\pa}{\pa \zeta^-}\right]^{\frac{M_2^2}{2i\nu}-\frac12}~
e^{ i \left( \nu \zeta^+ \zeta^- -  \zeta^+ p^- - \zeta^- p^+
+\frac{1}{\nu} p^+ p^- \right) }  \vert_{\zeta^\pm=0}
=(i\nu)^{\frac{M_1^2}{2i\nu}-\frac12}\\
\times \Gamma\left( \frac{M_1^2}{2i\nu} + \frac12 \right)
~\Gamma\left( \frac{M_2^2}{2i\nu} + \frac12 \right)
~(-p^+)^{\frac{M_2^2-M_1^2}{2i\nu}}
~U\left(  \frac{M_2^2}{2i\nu} + \frac12,~
1+\frac{M_2^2-M_1^2}{2i\nu};~\frac{i p^+ p^-}{\nu} \right)
\end{multline}
where $U(\lambda,\mu;z)$ is Tricomi's confluent hypergeometric function,
\be \label{Udef}
U(\lambda,\mu;z) = \frac{1}{\Gamma(\lambda)}~
\int_0^{\infty} dt~t^{\lambda-1}~(1+t)^{\mu-\lambda-1}~e^{-zt}
\ee
Reinstating the momentum-conserving delta functions along the
light-cone and the transverse directions, we  finally obtain
\begin{multline}
\langle 1 | :e^{i (p_2^+ X^- + p_2^- X^+) }:   | 3  \rangle =
g_s \, \delta_{-j_1 +j_2 + j_3}\, \delta\left(-\vec p_1+\vec  p_2  +
\vec  p_3 \right) ~
\exp\left[
- p_2^+ p_2^- \tDelta(\nu)  \right]\\
(2\nu)^{-1}~\left(-p_2^+\right)^{\mu-1} \left(-p_2^-\right)^{\tilde \mu-1}
U\left(\lambda, \mu, i \frac{p_2^+ p_2^-}{\nu}\right)
U\left(\tilde \lambda, \tilde \mu,i \frac{p_2^+ p_2^-}{\nu}\right)
\label{a3}
\end{multline}
where the cumulative effects of the zero and excited mode fluctuations
lead to a non-locality parameter
\be
\tDelta(\nu)\equiv\Delta(\nu)-\frac{1}{i\nu} = \psi(i\nu)+\psi(1-i\nu)-2\psi(1)
\ee
The parameters of the Whittaker functions appearing in \eqref{a3} are given by
\begin{align}
\label{lm}
\lambda &= \frac12+\frac{M_3^2}{2i\nu} &
\tilde\lambda &= \frac12+\frac{\tilde M_3^2}{2i\nu} \notag \\
\mu &= 1+\frac{M_3^2-M_1^2}{2i\nu} &
\tilde\mu &= 1+\frac{\tilde M_3^2-\tilde M_1^2}{2i\nu}\ .
\end{align}

In order to compute the real-space representation of the tachyon form
factor in the twisted sector, one needs to combine the excited mode
contribution \eqref{exctac} and the zero-mode form evaluated in the
previous subsection. The Fourier transform to
real space is most conveniently computed from the representation
\eqref{3ptz}, which leads to a simple Gaussian integral
\be
\label{3ptx}
\int  \langle f_1 | V_T(p_2) | f_3 \rangle e^{-i p_2 y}
=
\nu^2 ~{\cal D}_1^* ~{\cal D}_3 \cdot
e^{i \nu (\zeta^+ \zeta^-+\tzeta^+ \tzeta^-)
- \frac{1}{\tDelta(\nu)} (y^+ + \zeta^+ - \tzeta^+)(y^- + \zeta^-
- \tzeta^-) }
\vert_{\zeta^\pm=\tzeta^\pm=0}
\ee
Since this expression is still Gaussian in $\zeta$, the continuous
derivatives can be computed by employing \eqref{idw} as before. It is
interesting to notice that due to the zero-point fluctuations of
the  string in the twisted sector, the r.m.s. size of the
tachyon form factor is now given by
$\langle \delta y^+ \delta y^- \rangle \sim \tDelta(\nu)$.
In particular, for large winding
numbers, the twisted string appears to grow as $\sqrt{\log \nu}$.

We thus conclude that the three-point amplitude involving two twisted
strings and one untwisted, all in their ground state with respect to
transverse oscillations, is
finite, and proportional to the overlap of the zero-mode
wave-functions, up to a smearing due to the zero-point fluctuations
of the excited modes of the string.
Using \eqref{a3} as a source in the equation of motion of the untwisted
tachyon, we may further compute the back-reaction of a condensate
of winding strings on the tachyon field. We leave this to further work.

\subsubsection{Graviton form factor}
Let us now turn to the case where the untwisted state is a graviton.
Using the fact that only zero-modes $\a^\pm,\ta^\pm$
have a non-vanishing one-point function in the twisted vacuum, one
finds that the scattering amplitude of one graviton and two
twisted strings in their ground state is given by the overlap
\begin{multline}
\label{grav3}
\langle 1 | T^{\pm\pm} | 3 \rangle
=  \exp\left( - p_2^+ p_2^- \Delta(\nu) \right) \\
\langle f_1  |
\left(\ta^{\pm}  \mp i\nu p_2^\pm + \frac12 p_2^\pm  \right)
~e^{i (p_2^+ x^-+p_2^- x^+)} ~
\left(\a^{\pm} \pm i\nu p_2^\pm - \frac12 p_2^\pm  \right) |  f_3  \rangle
\end{multline}
In this expression, the two signs on $T^{\pm\pm}$ can be chosen
independently, and are correlated with the choices of sign for
$\a^\pm$ and $\ta^\pm$, respectively. The shift by
$\pm  i\nu p_2^\pm$ results from the definition of the normal ordered
vertex operator in \eqref{sha}. The shift by $\pm \frac12 p_2^\pm$
on the other hand results from commuting the zero-modes through
the plane wave exponential. This choice of ordering is particularly
appropriate for the real-space representation, where $\a^\pm$
are represented by covariant derivatives acting on the charged
wave function $f_3$, while  $\ta^\pm$
act as covariant derivatives on the wave function with opposite
charge $f_1^*$. Altogether, choosing $f_1=f_3$, we thus find that the graviton
one-point function in the twisted vacuum is given by the energy
momentum tensor of the complex wave function of the twisted string,
smeared under the action of the 
operator $\exp\left(  \Delta(\nu)  \pa^+ \pa^-\right)$.
This result is in line with the analogy to the charged particle
in a constant electric field discussed in \cite{Pioline:2003bs}.

The graviton one-point function for an on-shell twisted state is however
more easily computed in the oscillator representation, where
all $\a^+,\ta^-$ are moved to the right and $\a^-, \ta^+$
and to the left. It is tedious but easy to check that the same rule
as in \eqref{grav3} applies, namely that $\a^\pm$ and $\ta^\pm$
must be shifted by $\pm\frac12 p_2^\pm$ or $\mp\frac12 p_2^\pm$,
depending whether the operator is moved to the left or to the right,
respectively. For illustration purposes, we simply give
\begin{multline}
\langle 1 | T^{++} | 3 \rangle
= -(2\nu)^{-1}
\exp\left[ - p_2^+ p_2^- \Delta(\nu) \right] ~\left(-p_2^+\right)^{\mu} 
~\left(-p_2^-\right)^{\tilde \mu-2} \\
\left[ U(\tilde \lambda, \tilde \mu-1, \frac{i}{\nu} p_2^+ p_2^-)
- \frac12 (1-2 i\nu) p^+_2 p_2^-
U(\tilde \lambda, \tilde \mu, \frac{i}{\nu} p_2^+ p_2^-)\right] \\
\left[ U(\lambda-1, \mu+1, \frac{i}{\nu} p_2^+ p_2^-)
+ \frac12(1-2 i\nu) 
U(\lambda, \mu, \frac{i}{\nu} p_2^+ p_2^-)\right]
\end{multline}
From this result, we may now compute the back-reaction on the metric (and
dilaton, Kalb-Ramond fields) of a condensate of twisted strings, by simply
inserting $\langle 1 | T^{\mu\nu} | 3 \rangle$ into the equation of motion
of the graviton. In momentum space, this amounts to multiplying
\eqref{grav3} by the graviton propagator in De Donder gauge. We leave
the details of this computation to further work.

\subsection{Two-twisted two-untwisted scattering amplitude}
We now come to the scattering amplitude of two twisted (denoted 1,4) and
two untwisted (denoted 2,3) on-shell states. We  assume as before that the
twisted states are in their tachyonic ground state, while the
untwisted ones are either both tachyons.
By factorizing the amplitude in the limit of small momentum exchange,
one expects to recover the off-shell three-point functions computed
in the previous section.

By conformal invariance, one may fix the location of the
four vertex operators (1,2,3,4) at  $\infty, 1 , z, 0$, respectively,
and obtain the scattering amplitude by integrating over $z$:
\begin{multline}
\label{1234}
\langle 1 | T(2) T(3) | 4 \rangle
\equiv
\int dz\,d\bar z\,
\int_{-\infty}^{\infty} dv_2\,dv_3  ~e^{i(j_2 v_2 + j_3 v_3)}\\
\times  \langle 1 |: e^{i p_2^+(v_2) X^- + i p_2^-(v_2) X^+
+ p_2^i X^i }:~
:e^{i p_3^+(v_3) X^- + i p_3^-(v_3) X^+ + p_2^i X^i}:| 4 \rangle~
\end{multline}
The integral over $v_2,v_3$ projects the tachyon vertex operators
to orbifold-invariant states with  quantized momenta $j_2,j_3$.
As in \eqref{123}, the integral over $v_2+v_3$ (rendered compact
by the identification \eqref{boosta}) enforces
a momentum-conserving Kronecker symbol $\delta_{-j_1+j_2+j_3+j_4}$ which
we will reinstate
at the  end, allowing us to  set $v_2=0$ in the meantime.
The remaining integral over $v_3\equiv v$ however is non-trivial,
and is expected to be the  source of divergences, as in the
tree-level four-point
amplitudes of untwisted states investigated in Ref.~\cite{Craps:2002ii}.

Just as before, the light-cone contribution  to the second line of \eqref{1234}
can be computed by operator
methods, and factorizes into the product of a zero-mode piece and
an excited mode contribution:
\be
_{out}\!\langle f_1 |
e^{i p_2^+ X_0^- + i p_2^-  X_0^+ }
e^{i p_3^+(v) X_0^- + i p_3^-(v) X_0^+ }  | f_4 \rangle_{in}
\times\\
\langle \nu | e^{i p_2^+ X_{ex}^- + i p_2^- X_{ex}^+}
e^{i p_3^+ X_{ex}^- + i p_3^- X_{ex}^+} | \nu \rangle
\ee
Let us  start with the excited mode piece. As in the three-point
function \eqref{exctac},
the normal ordering of the tachyon vertex operators yields
a Gaussian factor $\exp[-(p_2^+ p_2^-+p_3^+ p_3^-)\Delta(\nu)]$.
In addition, one needs  to commute the positive frequency
part  of  $V_T(2)$ with the negative frequency part of $V_T(3)$,
leading to
\be
\exp\left(-[p_2^+ p_2^-+p_3^+ p_3^-]\Delta(\nu)
+2 \Re [ p_2^- p_3^+ e^{-v}  G(z;\nu)
+p_2^+ p_3^- e^{v} G(z;-\nu)]  \right)
\ee
where $G(z;\nu)$ denotes the propagator in the  twisted  sector
\be
G(z;\nu)\equiv
\sum_{n=1}^\infty \frac{z^{n+i\nu}}{n+ i\nu}
= \frac{z^{1+ i\nu}}{1+ i\nu}
~{}_2 F_1 \left(1+ i\nu, 1;2+ i\nu;z\right)
\ee
In the coincident limit  $z\to 1$,
\be
\label{coing}
G(z;\nu)\to
-\log(1-z)+\psi(1)-\psi(1+i\nu)
\ee
recovering the correct short distance singularity.

We now turn to the zero-mode piece. Using  the real-space
representation, it reduces to the overlap
\be
\label{r4}
\int dx^+ dx^- f_1^*(x^+,x^-)
e^{i (p_2^+ x^- + p_2^- x^+)}
e^{i p_3^+ e^{-v} x^-(z,\bar z) + i p_3^- e^{v} x^+(z,\bar z) }
f_4(x^+,x^-)
\ee
The latter can be computed by the now familiar generating function
technique,
\begin{multline}
\label{4ptz}
\nu^2 ~{\cal D}_1^* ~{\cal D}_4 \cdot
\exp\left[ i \nu (\zeta^+ \zeta^-+\tzeta^+ \tzeta^-)
- i \zeta^+ (  p_2^-+p_3^- e^{v} z^{-i\nu} )
-i \zeta^-  (p_2^++p_3^+  e^{-v} z^{i\nu} )  \right. \\  \left.
- i \tzeta^+ (p_2^-+p_3^- e^{v} \bar z^{i\nu} )
-i \tzeta^- (p_2^++p_3^+  e^{-v} \bar z^{-i\nu} )  \right. \\ \left.
+ \frac{i}{\nu}\left( p_2^+ p_2^ -+p_3^+ p_3^-
+ p_2^+ p_3^- z^{-i\nu}+ p_2^- p_3^+ \bar z^{-i\nu}
\right)  \right]
\vert_{\zeta^\pm=\tzeta^\pm=0}
\end{multline}
Evaluating  the continuous derivatives with the help of \eqref{idw},
and reinstating the momentum-conserving delta-function and the
standard contribution of the transverse coordinates $X^i$,
we finally obtain
\begin{multline}
\label{a4}
 \langle 1 |  T(2)  T(3) | 4 \rangle = g_s^2~ (2\nu)^{-1}
\delta_{-j_1 +j_2+j_3+j_4} ~\delta\left(-\vec p_1 +  \sum_{i=1}^3
\vec p_i\right)~\\
\int_{-\infty}^{\infty} \! dv  ~ e^{i j_3 v} ~\int dz d\bar z ~
\lvert 1- z \rvert^{2\vec p_2\cdot \vec p_3}
~\lvert z\rvert^{2\vec p_3   \cdot\vec p_4  +\vec p_3 \cdot \vec p_3 -2}
\exp\left[ - \left( p_2^+ p_2^-+p_3^+ p_3^- \right) \tDelta(\nu) \right] \\
\exp\left[
p_2^- p_3^+ e^{-v}
\left( G(z;\nu) +  G(\bar z;-\nu) + \frac{i}{\nu} \bar z^{-i\nu} \right)
+p_2^+ p_3^- e^{v}
\left( G(z;-\nu) +  G(\bar z;\nu) + \frac{i}{\nu} z^{-i\nu} \right)
\right]\\
\left[ 
p_2^+ +p_3^+ e^{-v} z^{i\nu} 
\right]^{\mu-1}
 \left[
p_2^-+p_3^- e^v {\bar z}^{i\nu}
\right]^{\tilde \mu-1} \,
U\left(\lambda, \mu,\frac{i}{\nu} A(v, z)\right) U\left(\tilde\lambda,\tilde
  \mu,\frac{i}{\nu}\tilde A(v, \bar z)\right)
\end{multline}
where $(\lambda,\nu,\tilde\lambda,\tilde\mu)$ are the same as in
\eqref{lm} after relabeling the particle $3$ into $4$, and
we defined
\bea
A(v,z)&=&(p_2^-+p_3^- e^{v} z^{-i\nu} )(p_2^++p_3^+  e^{-v}  z^{i\nu} ) \\
\tilde A(v,\bar z)&=&
(p_2^-+p_3^- e^{v} \bar z^{i\nu} )(p_2^++p_3^+  e^{-v}  \bar z^{-i\nu} )
\eea
Notice that
the correlation function in \eqref{a4} was evaluated for $|z|<1$, but
the same result holds when $|z|>1$ due to the transformation properties
of hypergeometric functions.

It is easy to check, using \eqref{coing},
that the amplitude in the limit  $z\to 1$ correctly factorizes
into the product of the three-point amplitude
of three untwisted fields with momenta $p_2^\pm, p_3^\pm,
p^\pm= p_2^\pm + e^{\mp v} p_3^\pm$, and
the amplitude $\langle 1 | T(p) |4 \rangle$ of two twisted
and one untwisted state, respectively:
\be
 \langle 1 |  T(2)  T(3) | 4 \rangle
= g_s~\int_{-\infty}^{\infty} dv~
 \langle 1 | ~ T(p^\pm= p_2^\pm + e^{\mp v} p_3^\pm;
j_2+j_3; \vec p_2+\vec p_3) ~  | 4 \rangle
\ee
where $ \langle 1 | T(p) | 4 \rangle$ is the off-shell three-point
function \eqref{a3}. One should in principle go to the basis
of boost eigenmodes for the intermediate untwisted state, but
we will refrain from doing so.

In the limit $z=0$ on the other hand, the amplitude should reduce to
a superposition of products of three-point functions $\langle 1 | T(2) |
M^2, \tM^2 \rangle$ and $\langle M^2, \tM^2 | T(3) |4 \rangle$,
where $|M^2,\tM^2 \rangle$ belongs to the continuum of two twisted
fields. In order to check that the amplitude
factorizes correctly, it important to note that as $z\to 0$, the limit
of $z^{\pm i\nu}$ is ill-defined. In view of the discussion
 in \cite{Berkooz:2004re},
we choose to rotate the world-sheet time slightly
away from the Euclidean axis, i.e. set $z=e^{i(e^{-i\theta} \tau - \sigma)}$
and $\bar z=e^{i(e^{-i\theta} \tau + \sigma)}$, respectively, with $\theta$
slightly smaller than $\pi/2$.
According to this prescription, as $\tau\to -\infty$
\be
\label{zinu}
z^{n+i\nu} \to 0 \quad \forall~n\geq 1\ ,\qquad
z^{n-i\nu} \to 0 \quad \forall~n\geq 0
\ee
The same holds upon replacing $z$ by $\bar z$, since both are related by
changing the sign of $\sigma$ ($z$ and $\bar z$ are therefore not complex
conjugate anymore !). In the limit $\tau\to -\infty$, the
third line of \eqref{a4} can therefore be dropped, leaving
simply the overlap of the zero-mode wave functions. By inserting a
completeness relation between states 2 and 3 in \eqref{r4}, the
correct factorization is therefore guaranteed. Nevertheless, it
is useful to compute the limit in some detail,
\begin{multline}
\label{a40}
 \langle 1 |  T(2)  T(3) | 4 \rangle  \to g_s^2~
\delta_{-j_1 +j_2+j_3+j_4} ~\delta\left(-\vec p_1 +  \sum_{i=1}^3
\vec p_i\right)~\\
\int_{-\infty}^{\infty} \! dv ~e^{i(j_3-j_1)v} ~
~ \int dz d\bar z ~
\lvert z\rvert^{2\vec p_3   \cdot\vec p_4 +
\vec p_3 \cdot \vec p_3 -2}
\exp\left[ - \left( p_2^+ p_2^-+p_3^+ p_3^- \right)
\tDelta(\nu)  \right] \\
(-1)^{\mu+\tilde\mu}~
(p_2^+)^{-\tilde\lambda}~
(p_2^-)^{-\lambda}~
(p_3^+)^{\mu-\lambda-1}~
(p_3^-)^{\tilde\mu-\tilde\lambda-1}~
z^{-\frac12 M_1^2 - \frac{i\nu}{2}}
\bar z^{-\frac12 \tM_1^2 - \frac{i\nu}{2}}
\end{multline}
where we used the asymptotics 
$U(\lambda,\mu,z) \sim z^{-\lambda}$ as $z\to \infty$.
Notice that, in the limit \eqref{zinu},
the only dependence in $v$ occurs via an
overall exponential factor, and therefore leads to a Dirac delta function
$\delta(j_3-j_1)$. Since the momenta $j$ are discrete variables,
this is strictly infinite whenever $j_3=j_1$ (and consequently
$j_2=-j_4$). Since on-shell twisted states in their tachyonic ground
state have $j_1=j_4=0$, this implies that the intermediate winding
state also has $j=0$. We therefore conclude that the four-point amplitude of 
two twisted states and
two untwisted states diverges\footnote{Strictly speaking, we cannot 
rule out a possible cancellation
of this singularity coming from another region of the domain of 
integration.} due to the propagation of winding strings with vanishing
boost momentum in the intermediate channel. 
This result closely parallels the discussion in
Ref.~\cite{Berkooz:2002je}, where tree-level scattering amplitudes
of 4 untwisted states were found to diverge, due to large graviton
exchange near the singularity. It would be interesting to see
whether the two divergences can somehow be canceled, or
whether quantum corrections, for example in the eikonal approximation,
lead to a finite amplitude.

\section{Scattering amplitudes with more than two twisted strings}
In the previous section, we have computed scattering amplitudes
involving exactly two twisted states, as well an arbitrary number
of untwisted states, using Hamiltonian  quantization in the twisted vacuum.
For more twisted states however, operator techniques are no longer
suited, and one should in principle resort to twisted vertex
operators. Unfortunately, the conformal theory of twisted vertex
operators with a continuous spectrum has not been developed  yet.
Nevertheless, we  shall be able to obtain scattering amplitudes
involving three  or more twisted states by mapping our problem to an
analogue one which has already  been solved: the Wess-Zumino-Witten
model of a four-dimensional plane wave \cite{Nappi:1993ie,Olive:1993hk}.

\subsection{Misner space and the Nappi-Witten plane wave}
The WZW model based on the non-semi-simple group $H_4$, i.e. the
three dimensional  Heisenberg  algebra $[P,\bar  P]=K$  extended
by the generator of rotations in the complex  $P$ plane, is known
to describe a four-dimensional plane wave, with metric and Neveu-Schwarz
flux
\be
\label{ds}
ds^2  = -2 du dv + d\zeta d\tzeta - \frac14 \zeta\tzeta du^2\  ,\qquad
H= du dx  d\bar x
\ee
where $\zeta=x_1+ix_2$  is the complex coordinate in the plane.
In the light-cone gauge $u=p \tau$, it is known that the transverse
coordinate $X$ has the mode expansion of a complex scalar field twisted by  a
real angle proportional to the light-cone
momentum $p$~\cite{Kiritsis:jk}. In fact,
there exists a free-field representation where the vertex operator
of a physical state with non-zero $p$ is just the  product
of  a plane wave along the $(u,v)$ light cone coordinates,
times a twist field\footnote{For integer $p$, new ``spectrally flowed''
states appear describing long strings stabilized by the
flux \cite{Kiritsis:2002kz}.}
creating a cut $z^p$ on the world-sheet,
where $p$ is in general an irrational
number\footnote{Orbifolds with irrational angle have been recently
discussed in \cite{Kutasov:2004aj}.}.
Correlation functions of physical states
have been computed using standard WZW techniques
\cite{D'Appollonio:2003dr,Cheung:2003ym},
and, by removing the plane wave contribution, it is then possible
to extract the correlator of twist fields with arbitrary angle.

For our purposes, it is more convenient to use the current algebra
techniques of \cite{D'Appollonio:2003dr}\footnote{It  is also possible
to  use a Wakimoto-like  free-field representation to compute
scattering amplitudes of an arbitrary number of twiste fields
\cite{Cheung:2003ym,Bianchi:2004vf}.}. These authors introduce a semi-classical
representation for states  with non-zero $p$ as
\be
\label{vp}
\Phi^\pm_{p,q}=\exp\left(
\mp i p  v +  i q u \right)\cdot
\exp\left( - \frac{p}{2} \zeta  \tzeta
+ i\,p\,\zeta\,x  e^{\pm i  u/2}
+ i\,p\,\tzeta\, \bar x  e^{\pm i  u/2}
+ p\,x \bar x e^{\pm  i u}
\right)
\ee
where $u,v,\zeta,\tzeta$ are the coordinates in the plane
wave geometry \eqref{ds}, and the variables $x,\bar x$
parameterize a representation $V^\pm_{p,j}$ of $H_4$,
with central element $K=ip$ and quadratic Casimir $C=-2p(j\pm 1/2)$.
In particular, by expanding in arbitrary powers of $(x,\bar x)$
one obtains the whole tower of states on which the
isometry group $H_4$ acts:  these states are most conveniently
understood  as excitations of a two-dimensional harmonic oscillator
with frequency $p$. In  the conformal field  theory,
the variables $u,v,\zeta,\tzeta$ become fields, and Khnizhnik-
Zamolodchikhov  equations relate world-sheet derivatives
with derivatives with respect to $x,\bar  x, j, v$, allowing to
solve for the conformal blocks.

Our main observation is that the semi-classical vertex \eqref{vp}
closely resembles the off-shell  wave  functions
\eqref{phiin},\eqref{phiout}  which we introduced to describe
twisted strings in the Misner geometry, in particular  the generating
coordinates $x,\bar x$ describing the tower of harmonic oscillator
states are quite similar  to the continuous generating coordinates
$\zeta^\pm,\tzeta^\pm$ describing the continuum of states in the
inverted harmonic oscillator. To  make  the connection
more precise, notice that the  {\it in} and {\it out} wave functions can be
written in the $(\tzeta^+,\zeta^-)$ and $(\zeta^+,\tzeta^-)$ representations
respectively as
\bea
f(x^+,x^-) &=& {\cal D}^{in} \cdot
\exp\left( - i\frac{\nu}{2} x^+ x^-
-i\nu  \tzeta^+ x^-  -i\nu  \zeta^- x^+
- i\nu \tzeta^+ \zeta^-
\right) \\
f^*(x^+,x^-) &=& ( {\cal D}^{out} )^* \cdot
\exp\left( - i\frac{\nu}{2} x^+ x^+
-i\nu  \zeta^+ x^-  -i\nu  \tzeta^- x^+
-i\nu \zeta^+ \tzeta^-
\right)
\eea
This is equal to the second factor in $\Phi^{\pm}_{p,q}$,
respectively, upon identifying
\begin{gather}
\label{vpz}
p=i\nu\ ,\quad
x^-=i\tzeta\ ,\quad
x^+=-i \zeta \ ,\quad\\
\zeta^-=x ~e^{iu/2}\  ,\quad
\tzeta^+=-\bar x ~e^{iu/2}\ ,\quad
\tzeta^-=\bar x~e^{-iu/2}\ ,\quad
\zeta^+=-x~ e^{-iu/2}
\end{gather}
Finally, the $p=0$  vertex operators
\be
\Phi_{s,j}^0 = \delta(x\bar x e^{iu})~
\exp\left(iju\right) \times \exp\left[    \frac{i}{\sqrt{2}} s
\left(\zeta  \sqrt{\frac{x}{\bar x}} \
+ \tzeta \sqrt{\frac{\bar x}{x}} \right) \right]
\ee
agree with the off-shell untwisted  wave functions
$\exp(i  p^+ x^- + i p^- x^+)$ upon identifying
\be
\label{v0z}
p^- =- \frac{i  s}{\sqrt{2}} \sqrt{\frac{x}{\bar x}} \  ,\quad
p^+ =\frac{i  s}{\sqrt{2}} \sqrt{\frac{\bar x}{x}}
\ee
Having identified the parameters in the plane wave in
terms of those in the Misner space, it is important to stress
some  differences between the two  computations:
\begin{itemize}
\item The plane wave computations  involves a path integral over the
$u$ and $v$ variables, in particular over their  respective
  zero-modes. The  integral over $v_0$ simply implies conservation
of the light-cone momentum $p$, i.e. conservation of the winding
number $w$. The integral over $u_0$ instead amounts to Fourier
transforming the Misner amplitude under the action
\be
\zeta^\pm  \to e^{\mp iu/2} \zeta^\pm \ ,\quad
\tzeta^\pm  \to e^{\pm iu/2} \tzeta^\pm
\ee
which therefore needs to be undone if one is to map plane wave
amplitude to Misner space.
\item In the plane wave background, setting $x=\bar x=0$ gives
scattering amplitudes for the ground state  of the harmonic
oscillator. This is  not so for Misner amplitudes, which instead
should be obtained by applying the continuous derivative operator
\eqref{difdef} for arbitrary (complex)  order, or equivalently
Eq. \eqref{idw}. This is the analytic continuation
prescription which allows us to obtain physical  states of the
inverted harmonic oscillator from states of the standard harmonic
oscillator.
\item The plane wave amplitudes do {\it not} involve a projection
over boost invariant states. While twisted  states, by virtue of the
continuous derivation prescription, automatically have a
well-defined boost momentum, untwisted  vertex operators need
to be smeared under the action of the boost.
\item Due to the analytic continuation $p\to i\nu$, the analytic behavior
of the Misner amplitudes as a function of $\nu$ is quite different from
the plane wave case. For example, the logarithmic growth
of the non-locality parameter $\Delta(\nu)$ introduced
in \eqref{exctac} is replaced by a set of poles at integer $p$
corresponding to the production threshold of long
strings \cite{D'Appollonio:2003dr}.
\end{itemize}
With these words of caution, it is easy to see that the two-point
\eqref{2ptz}, three-point \eqref{3ptz} and four-point \eqref{4ptz}
functions of twisted states in Misner space match the plane wave
result.

\subsection{Three-twist amplitude}
We now wish to extract the three-twist amplitude from the plane wave
result \cite{D'Appollonio:2003dr,Cheung:2003ym},
\be
\label{kd3}
\langle 1,2| 3 \rangle
=\delta_{\mathbb{N}}(L)~\delta(p_3-p_2-p_1)~
\left| e^{-x_3(p_1 x_1+p_2 x_2) }(x_2-x_1)^{L} \right|^2
\times \frac{1}{\Gamma(1+L)}
\left[ \frac{\gamma(p_3) }{\gamma(p_1) \gamma(p_2)} \right]^L
\ee
where $L=q_3-q_1-q_2$ and $\gamma(p)\equiv \Gamma(p)/\Gamma(1-p)$.
In this expression, the first terms originate purely from kinematics,
while the last term is the OPE coefficient, obtained from
factorization
of the four-point amplitude\footnote{The $1/2$ in
the exponent \cite{D'Appollonio:2003dr}, eq. 4.0.27 can be reabsorbed in
the normalization of the vertices \cite{Bianchi:2004vf}.}.
It is  interesting to note that the OPE coefficient emerges naturally
in the Wakimoto representation as a Virasoro-Shapiro-type
integral \cite{Cheung:2003ym},
\be
\frac{1}{2\pi}\int d^2 w~|w|^{2(p_1-1)} |1-w|^{2(p_2-1)}
=  \frac{\gamma(p_1) \gamma(p_2)}{\gamma(p_1+p_2)}
\ee
hence inherits the usual Regge and Gross-Mende behavior as a  function
of the Mandelstam-like  variables $(s,t,u)\sim(p_1,p_2,p_3)$.

Our first task is to reproduce the kinematics from the overlap
of three twisted off-shell wave functions in the
$\zeta^\pm,\tzeta^\pm$
representation:
\be
\label{ov3}
\int dx^+  dx^- [ f_1 f_2 ]^* f_3
= e^{-i\zeta_3^- (\nu_1 \zeta_1^+ + \nu_2 \zeta_2^+)
     -i\tzeta_3^+ (\nu_1 \tzeta_1^- + \nu_2 \tzeta_2^-)}
e^{-\frac{i\nu_1 \nu_2}{\nu_3}
\left(\zeta_1^+ - \zeta_2^+ \right)
\left(\tzeta_1^- - \tzeta_2^- \right)}
\ee
with $\nu_1+\nu_2=\nu_3$. After substituting the plane wave variables,
only the last term in the exponential becomes $u$-dependent;
the integral over the  zero-mode of $u$ leads to
\be
\int_{-\infty}^{\infty} du_0 ~ e^{-i L  u_0} ~
e^{-\frac{i\nu_1 \nu_2}{\nu_3 }
\left(\zeta_1^+ - \zeta_2^+ \right)
\left(\tzeta_1^- - \tzeta_2^- \right) e^{iu_0}}
=\frac{1}{\Gamma(1+L)}
\left[
\frac{\nu_1 \nu_2}{\nu_3 }
\left(\zeta_1^+ - \zeta_2^+ \right)
\left(\tzeta_1^- - \tzeta_2^- \right) \right]^L
\ee
This reproduces the $(x_2-x_1)^L$ term in \eqref{kd3}, as  well
as the small $p$ behavior  of the OPE coefficient, since
\be
\frac{\gamma(p_3) }{\gamma(p_1) \gamma(p_2)} \sim
\frac{p_1 p_2}{p_3}
\ee
Reversing the argument, we find that the complete 3-twist
amplitude in Misner space is given by the generating function
\begin{multline}
\langle 1,2| 3 \rangle =
(\nu_1\nu_2\nu_3)~
\delta(\nu_1+\nu_2-\nu_3)~ {\cal D}_1^*{\cal D}_2^*{\cal D}_3  \cdot
\exp\left[ -i\zeta_3^- (\nu_1 \zeta_1^+ + \nu_2 \zeta_2^+)
     -i\tzeta_3^+ (\nu_1 \tzeta_1^- + \nu_2 \tzeta_2^-) \right.\\
\left. +\frac{\gamma(i \nu_3) }{\gamma(i\nu_1) \gamma(i\nu_2)} ~
\left(\zeta_1^+ - \zeta_2^+ \right)
\left(\tzeta_1^- - \tzeta_2^- \right) \right]
\end{multline}
We can now revert to the real space representation, and find that,
instead of taking place at a single point $x_1^\pm=x_2^\pm=x_3^\pm$
as in \eqref{ov3}, the interaction is in fact spread with a kernel
\be
\label{ker3}
\int dx^\pm_1  dx^\pm_2~ [ f_1(x_1^\pm) f_2(x_2^\pm) ]^*
~\exp\left[ (x_1^+ - x_2^+)(x_1^- - x_2^-) \Xi(\nu_1,\nu_2) \right]
~f_3\left( x_3^\pm - \frac{\nu_1 x_1^\pm + \nu_2 x_2^\pm}{\nu_1+\nu_2}
\right)
\ee
where the size of the non-locality is given by the (real) ratio
\be
\label{xi}
\Xi(\nu_1,\nu_2) = -i \frac{
1- \frac{i \nu_3}{\nu_1\nu_2}
\frac{ \gamma(i \nu_3)}{ \gamma(i \nu_1) \gamma(i \nu_2)}}
{1+ \frac{i \nu_3}{\nu_1\nu_2}
\frac{ \gamma(i \nu_3)}{ \gamma(i \nu_1) \gamma(i \nu_2)}}
\ee
As $\nu_i\to 0$, $\Xi(\nu_1,\nu_2) \sim 1/ (2\zeta(3)~\nu_3^2)$
so that the interaction becomes local. For larger $\nu$ however,
the non-locality scale $1/\sqrt\Xi$ diverges when
$\nu_1 \nu_2 \gamma(i \nu_1) \gamma(i \nu_2) =
i\nu_3 \gamma(i \nu_3)$ , before switching sign and vanishing
again when $\nu_1 \nu_2 \gamma(i \nu_1) \gamma(i \nu_2) = -
i\nu_3 \gamma(i \nu_3)$ (e.g, for $\nu_1=\nu_2$ this is found
numerically to happen
at $\nu_1\sim 1.236$ and $\nu\sim 3.039$, respectively).
It would be interesting to understand the physical origin of this diverging
non-locality scale.

We conclude that scattering amplitudes involving 3 or more twist fields
can be computed from scattering amplitudes in the Nappi-Witten plane
wave by a simple analytic continuation (and a possibly large number of
Fourier transforms). It would be interesting to develop the conformal
field theory of continuous Lorentzian twist fields, and obtain a Wakimoto-type
representation for computing scattering amplitudes including both
twisted and untwisted states.
This result opens the way to compute the
back-reaction of a condensate of twisted states on the twisted fields
themselves: deforming the world-sheet action with $\lambda V_{w} +
\bar\lambda V_{-w}$, it is necessary to include a deformation
by $V_{\pm 2w}$ at order $|\lambda|^2$ to preserve conformal invariance.
We leave a more detailed analysis for future work.

\vspace*{1cm}

\noindent {\it  Acknowledgments}: The  authors are happy to thank
G. d'Appollonio, Y. Cheung, E. Kiritsis, D. Kutasov, G. Moore, E. Rabinovici, 
M. Rozali, K. Savvidy for
useful discussions. B.P. is grateful to the Weizmann Institute
for hospitality during part of this project, and to Rochelle and
Phalevi for their patience during the last stages of this research.
M.B. is grateful to Ecole Polytechnique and to the 
Erwin Schr\"odinger Institute for their hospitality during part of
this work. The work of M.B. is supported by Israeli
Academy of Science centers of excellence program,
by the Minerva Foundation, by EEC
RTN-2000-001122 and by the Einstein Center for Theoretical Physics. 

\appendix

\vskip 1cm
\centerline{\large \bf Appendix}
\vskip 1cm

\section{Scattering amplitudes of untwisted strings}
Tree-level scattering amplitudes for untwisted
states for string theory in Misner space have been investigated 
in \cite{Berkooz:2002je}, revealing divergences
from a kinematical regime governed by the Regge limit of the
standard Virasoro-Shapiro amplitude. The divergences can be 
traced back to large graviton exchange near the singularity, as the
particles experience a high blue-shfit. The authors of \cite{lms2} have
demonstrated that similar divergences in the parabolic orbifold
can be regularized by deforming the background to the null-brane
space-time \cite{Figueroa-O'Farrill:2001nx,Simon:2002ma}
(i.e. combining the null boost with a translation, much
as  Grant space is a deformation of the
Misner space): in this deformed background, wave-packets with a 
continuous momentum distribution, except for remaining collinear
divergences \cite{lms2}.  In this Appendix, we perform a similar analysis for 
tree-level scattering amplitudes of untwisted states in Grant space. We
find that, unlike the null-brane case, divergences remain even for 
non-zero momentum transfer, although the range of dangerous momenta
is reduced with respect to Misner space.

\subsection{Four-graviton amplitude in type II superstring}\label{GRANTsec1}
In order to avoid tachyon-related poles, we consider the type II superstring
on Grant space, i.e. the orbifold of Minkowski space $\Real^{1,2}$
by the orbifold action
\begin{equation}
\label{grantdef}
(x^+,x^-,x_2)\rightarrow (e^{2\pi\beta} x^+,e^{-2\pi\beta}x^-,x_2+2\pi R)
\end{equation}
The vertex operator for NS-NS massless states is given, in the $(-1,-1)$
superghost picture, by
\begin{equation}\label{untwwave}
\V^{-1,-1}= g_s
\int d^2z\int_{-\infty}^\infty d u\,\zeta_{\mu\nu}\cno{e^{-\phi-\tilde{\phi}}\, \psi^\mu \tilde{\psi}^\nu\,e^{iP\cdot X}}
\end{equation}
where 
$P=\left(e^{-u}\ppp,e^{u}\ppm,p,\vec
p\right)$ is the momentum on the covering space, and 
$~\zeta_{\mu\nu}$ a polarization tensor (which also depends on 
$j,u$). Invariance under the orbifold group imposes the
condition
\begin{equation}
    p=\frac{n-\beta j}{R}\csp1 n\in\Zf \csp1 p,j\in \Rf.
\end{equation}
We also use the notations of \cite{Berkooz:2002je} for the
two-dimensional mass:
\begin{equation}
    m^2\equiv2\ppm\ppp=\vecs p+p^2
\end{equation}
As in \cite{Berkooz:2002je}, the tree-level scattering amplitude of untwisted 
states is obtained by smearing the standard
scattering amplitude for four massless NS-NS fiels in flat space
\cite{Schwarz:1982jn} over the action of the boosts,
\begin{multline}\label{untw4nsns}
    \A_{4NS}=-{i\pi^2g_s\at^3}R^4\tp^{11}\delta^{7}\left(\sum_i\vec p_i\right)\delta\left(\sum_in_i\right)\delta\left(\sum_ij_i\right)\cdot\\
    \cdot\int_0^\infty dv_4 \,\hat K\, H(s) H(t) H(u) \frac{v_{2}^{-ij_2+1} v_{3}^{-ij_3+1} v_4^{-ij_4-1}}
    {\abs{p^+_2 p^-_3 v_{3}^2 - p^+_3 p^-_2 v_{2}^2}},
\end{multline}
where $s,t,u$ denote the Mandelstam variables of the boosted particles, e.g.
\be
    s=2\left(p^+_1 p^-_2 v_{2} + p^-_1 p^+_2\frac1{v_{2}}\right) - 2p_1 p_2 -
    2\vecd{p_1}{p_2} \ ,
\ee
the function $H(x)=
={\Gamma\left(-\frac{\at}4 x \right)}/{\Gamma\left(1+\frac{\at}4 x \right)}$
and the kinematical factor
\begin{align}
    \hat K=&t^{\mu _1\ldots \mu_4}{}_{\nu1\ldots\nu_4} t^{\rho_1\ldots \rho_4}{}_{\s_1\ldots\s_4} \prod_{r=1}^4\zeta^{r}_{\mu_r\rho_r}
    P_{r}^{\nu_r}P_{r}^{\sigma_r}\ .
\end{align}
In this expression, the $P_{i}$'s are the covering-space momenta, and the 
$t^8$ tensors can be found in \cite{Schwarz:1982jn}. Using momentum
conservation on the covering space, two of the variables
variables ($v_2$ , $v_3$) have been written as the positive solution(s) of the
following system\footnote{The $v_i$'s are defined from the
original $u_i$'s by the change of variables: $~ v_i\equiv e^{u_1-u_i}$.}:
\begin{equation}\label{untwsysofeq}
\begin{cases} p_1^- + p_2^- v_2 + p_3^- v_3 + p_4^- v_4 &= 0 ,\\
              p_1^+ + p_2^+\frac{1}{v_2} + p_3^+\frac{1}{v_3} + p_4^+\frac{1}{v_4} &= 0 .\\
\end{cases}
\end{equation}
The limit of $~v_4\rightarrow\infty~$ is the familiar Regge
limit\footnote{Note that the $~v_4\rightarrow0~$ limit gives
similar results when evaluated to leading order in $~1/v_4~$,
switching the role of  $~p^+_i\leftrightarrow p^-_i~$ and
$~v_i\leftrightarrow1/v_i~$.}. The solutions for $v_2$ , $v_3$ to
the leading order in $v_4$ are:
\begin{gather}
    p^+_4p^+_2<0\csp1p^+_3p^+_1<0\rar1v_2=-\frac{p^-_4}{p^-_2}v_4\csp1v_3=-\frac{p^+_3}{p^+_1}\cr
    p^+_4p^+_3<0\csp1p^+_2p^+_1<0\rar1v_2=-\frac{p^+_2}{p^+_1}\csp1v_3=-\frac{p^-_4}{p^-_3}v_4
\end{gather}
We focus on the first solution where $~s\sim-2p^+_1p^-_4 v_4~$ is
large and $t$ is finite. The contribution to the amplitude from this
range is:
\begin{multline}\label{untw4nsReg}
    \A_{4NS}\sim-\frac{i\pi^2g_s\at^4}{2}R^4\tp^{11}\delta^{7}\left(\sum_i\vec p_i\right)\delta\left(\sum_in_i\right)\delta\left(\sum_ij_i\right)\hat K_4\\
    \times ~ \int_{\Lambda\gg1}^\infty d\s\,
    \left(-\frac{p^-_4}{p^-_2}\right)^{-ij_2}\left(-\frac{p^+_3}{p^+_1}\right)^{-ij_3}
    \abs{-\frac\at2p^+_1p^-_4}^{1+\frac{\at t}2}
    \frac{\Gamma\left(-\frac{\at t}4\right)}{\Gamma\left(1+\frac{\at t}{4}\right)}
    e^{\left(1+\frac{\at t}2+ij_1+ij_3\right)\s}\\
\end{multline}
where $\hat K_4$ is the leading term in $\hat K$ as $v_4\to\infty$,
\begin{equation}
    \hat K=\hat K_{4}v_4^4+O(v_4^3).
\end{equation}
For the superstring case $t$ is strictly negative, so we find
divergences within the kinematical regime:
\begin{equation}\label{untw4nsunsmconv}
    -t=\left(\vec{p}_1+\vec{p}_3\right)^2 +
    \left(\frac{(n_1+n_3)-\beta (j_1+j_3)}R\right)^2 \leqslant \frac2{\at}
\end{equation}
generalizing the result obtained for the bosonic string in Misner space
in \cite{Berkooz:2002je}.

\subsection{Scattering amplitudes for wave-packets in Grant space}
\label{secREGofDIV}

As an attempt to remove the divergences found in the four-point
amplitudes we follow the procedure of \cite{lms2}, and consider
wave-packets constructed out of a superposition of 
wave-functions with different momenta $j$. 
Let us start from wave functions with a well-defined value of
$(j,\vec p)$ \cite{Nekrasov:2002kf}:
\begin{equation}\label{untwJKMbase}
    \Psi_{n,j,m,\epsilon^\pm,\vec p\,}(\Xp,\Xm,X,\vec X)=\sqrt{\frac{m}{2\pi R}}\, e^{i\vecd pX+i\frac{n-\beta j}{R}X}\int_{-\infty}^\infty du\, e^{-i\epm\frac{m}{\sqrt2}\Xp e^u-i\epp\frac{m}{\sqrt2}\Xm
    e^{-u}-iju},
\end{equation}
where $~ m=\sqrt{2\abs{\ppp\ppm}}>0 ~$ and $~(\epp)^2=(\epm)^2=1$.
The wave-functions (\ref{untwJKMbase}) form a complete basis
in Grant space, however they are badly behaved near the light-cone
($\Xp\Xm=0$):
\begin{equation}
    \Psi_{n,j,m,\epsilon^\pm,\vec p\,}
    \sim
    \left[\frac{1-\coth{\pi j}}{\Gamma(1+ij)}\left(\frac{m}{\sqrt{2}}\abs{\Xp}\right)^{i j}
    -\frac{\csch{\pi j}}{\Gamma(1-ij)}\left(\frac{m}{\sqrt{2}}\abs{\Xp}\right)^{-i j}\right]+O(\Xp,\Xm)
\end{equation}
In the latter formula we chose explicitly $\epp\Xm,\epm\Xp>0$.
The wave functions for which we can expect the best behavior are the ones that
vanish near the light cone. For this, we consider wave-packets
\be
\int dj ~F(j)~    \Psi_{n,j,m,\epsilon^\pm,\vec p\,}
\ee
such that
\begin{equation}\label{untwcond}
    \int_{-\infty}^\infty dj F(j) x^{ij}=\tilde F(\log x)\xrightarrow{x\rightarrow0}0,
\end{equation}
where $\tilde F$ is the Fourier transform of $F$. In the Regge limit,
 the four-point amplitude for such wave packets is given by
\begin{multline}
    \prod_{i=1}^4\left(\int_{-\infty}^\infty dj_i F^{(i)}(j_i)\right)\A_{4NS}(j_1,j_2,j_3,j_4)=\\
    =\int dj_1dj_2djd\tilde j\,F^{(1)}(j_1)F^{(2)}(j_2)F^{(3)}(j-j_1)F^{(4)}(\tilde j-j_2)\,\A_{4NS}(j_1,j_2,j-j_1,\tilde j-j_2)
\end{multline}
The integration over $j$ amounts to a Fourier transform evaluated at
$~\s\rightarrow\infty~$. Thus the amplitude can be evaluated
as\footnote{With the hindsight that the divergences arise from
$\abs{\at t}\leq2$, we expand the $\Gamma$ functions near $~t=0$.}:
\begin{align}\label{untwsmearj}
    \prod_{i=1}^4\left(\int_{-\infty}^\infty dj_i F^{(i)}(j_i)\right)&\A_{4NS}(j_1,j_2,j_3,j_4)=\cr
    &\hsp{-5}\delta^{(7)}\Big(\sum_ip_i\Big)\delta\Big(\sum_in_i\Big)\hat
    K_4\abs{-\frac\at2p^+_1p^-_4}^{1-\frac\at2\vecs p_t}\int_{\Lambda\gg1}^\infty d\s\,e^{\s\left[1-\frac\at2\vecs p_t\right]}
    \cr
    &\hsp{-5}\Biggl[\left(\widetilde F^{(2)}\widetilde F^{(4)}(x)\right)
    \ast
    \left(\widetilde F^{(1)}\widetilde     F^{(3)}(x)\right)\ast\left(\frac1{\sqrt\s}e^{-\frac{x(x-i2\at\frac{n}{R}E\s)}{4E^2\s}}\right)\ast\left(e^{i\frac{n}{\beta}x-\abs{x}\sqrt{\frac{\vecs p_t}{E^2}}}\right)
    \Biggr]_{x=\s}\cr
\end{align}
where $~E=\beta/ R\csp{0.5}\vec
p_t=p_1+p_3\csp{0.5}n=n_1+n_3~$, and $\ast$ denotes the convolution product
(i.e. multiplication of the Fourier transforms).

At large $\s$ the convolution is dominated by the slowest decaying
function. If all the Fourier transforms $\widetilde F^{(i)}(x)$ go
to zero  at $~\abs{x}\rightarrow\infty~$ faster then
$~e^{-|x|\sqrt{{\vecs p_t}/{E^2}}}~$ the amplitude
(\ref{untwsmearj}) behaves at the Regge limit as:
\begin{equation}
    \sim\delta^{(7)}\Big(\sum_ip_i\Big)\delta\Big(\sum_in_i\Big)\hat
    K_4\abs{-\frac\at2p^+_1p^-_4}^{1-\frac\at2\vecs p_t}\int_{\Lambda\gg1}^\infty d\s\,\exp\left[\s\left(1-\frac\at2\vecs p_t-\sqrt{\frac{\vecs
    p_t}{E^2}}\right)\right].
\end{equation}
Alas, the range of divergences from (\ref{untw4nsunsmconv})
is reduced but not removed:
\begin{equation}\label{smearond}
\left(\vec{p}_1+\vec{p}_3\right)^2
\leq \frac{(\sqrt{1+2\at E^2}-1)^2}{(\at   E)^2}\ ,\qquad E=\frac{\beta}{R}
\end{equation}
As $R\to 0$, this reduces to \eqref{untw4nsunsmconv} as expected.

\end{document}